\documentclass[preprint2]{aastex}

\def\lesssim{\mathrel{\hbox{\rlap{\hbox{\lower4pt\hbox{$\sim$}}}\hbox{$<$}}}}
\def\gtrsim{\mathrel{\hbox{\rlap{\hbox{\lower4pt\hbox{$\sim$}}}\hbox{$>$}}}}
\newcommand{\beq}{\begin{equation}}
\newcommand{\eeq}{\end{equation}}
\newcommand{\beqa}{\begin{eqnarray}}
\newcommand{\eeqa}{\end{eqnarray}}

\usepackage{graphicx}

\begin{document}

\title{Life Products of Stars}
\author{Aldo M. Serenelli}
\affil{Institute for Advanced Study, Einstein Drive, Princeton, NJ 08540, USA}
\and
\author{Masataka Fukugita}
\affil{Institute for Advanced Study, Einstein Drive, Princeton, NJ 08540, USA}
\affil{Institute  for  Cosmic  Ray  Research, University  of  Tokyo,
  Kashiwa 277-8582, Japan}

\begin{abstract}
We attempt to document complete energetic transactions of stars in
their life.  We calculate photon and neutrino energies that are
produced from stars in their each phase of evolution from 1 to 8${\rm
M}_\odot$, using the state-of-the-art stellar evolution code, tracing
the evolution continuously from pre-main sequence gravitational
contraction to white dwarfs. We also catalogue gravitational and
thermal energies and helium, and heavier elements that are stored in
stars and those ejected into interstellar space in each evolutionary
phase.
\end{abstract}


\maketitle

\section{Introduction}

Stars are taken as an engine that uses hydrogen as a fuel, producing
energy mostly in photons and leaving helium and heavier elements as
ashes. Baryons in stars are only 6\% of those in the Universe, yet
stars cause varieties of most spectacular phenomena, and also give a
key to understanding the energetics of the Universe (\citealp{fp04}; hereafter FP).
FP presented an accounting of energies
that are relevant to stars using rather rough estimates in view of the
difficulty in assembling the relevant astrophysical quantities. There
are many calculations of stellar evolution done to date, some of the
most recent examples being the Geneva tracks \citep{gen92,char96} 
 and the Padova tracks \citep{gir00}
among others, but quantities like integrated luminosities and
gravitational binding energies stored in stars are 
not explicitly presented.

The purpose of this paper is to present evolution of stars mainly from
the energetics point of view. For example, how much radiation energy
is produced in main sequence phase, how much in red giant branch and
asymptotic branch phases, and how much energy is stored in heavy
elements as well as in gravitational binding, and so forth? The energies in
radiation observed in the extragalactic background light must balance
those produced by stars besides a small fraction contributed by
active galactic nuclei, and they are documented in the nuclear binding
energy as a fossil record.  
We also want to know how much fraction of helium and heavy element
produced in stars are liberated into interstellar space.
Then, what fraction of energies go to
neutrinos, which account for the difference between the nuclear binding
energy and the energy in radiation? 

We focus our calculations on low and intermediate mass stars from 1 to
8 solar masses that eventually end up as white dwarfs, and we follow
the evolution from pre-main sequence contraction to the
cool white dwarf with luminosity $\log(L/L_\odot)=-4.5$. This enables
us to compute integrated output of stars in a consistent way.  We also
add the case for 0.8${\rm M}_\odot$, calculated to the beginning of
helium ignition to allow a continuation to lower mass stars.
We consider stars with solar metallicity.

We remark that white dwarfs are the most important reservoir of heavy
elements today, which stores about 80\% of heavy element in the
Universe. The gravitational binding energy stored in white dwarfs is 5
times that in main sequence stars in the cosmic mean. Stellar winds in
the thermal-pulse (TP) AGB phase are also the important source of
chemical enrichment besides supernovae. Thus, white dwarfs are one of
the most prominent fossil records of energy transactions in the
Universe (FP).

\section{Calculations}

  Calculations have been carried out with the LPCODE stellar evolution
code \citep{alt03}. 
For this work, we have adopted an extended nuclear network
involving 32 isotopes from $^1$H to $^{32}$P linked by 96 reactions
that cover all relevant nuclear processes from detailed hydrogen and
helium burning up to a simplified yet energetically accurate
treatment of carbon and neon burning. The NACRE compilation
\citep{nacre} gives the basic set of nuclear reaction rates adopted in
this work. The two key reactions $^{14}{\rm N(p,\gamma)^{15}O}$ and
$^{12}{\rm C(\alpha,\gamma)^{16}O}$ are from \citet{luna04} and
\citet{kunz02} respectively.  Carbon and neon burning rates are taken
 from \citet{cf88}.  In our calculations, carbon burning is relevant
only to the 8 M$_\odot$ model.

The equation of state used from the pre-main-sequence up to the
beginning of the white dwarf cooling branch includes, for the
low-density regime, partial ionization for hydrogen and helium
compositions, radiation pressure and ionic contributions. For the
high-density regime, partial electron degeneracy and Coulomb
interactions are also included. For the white dwarf regime we adopt in 
regions of partial ionization an
updated version of the equation of state of \citet{mm79} that 
accounts for covolume, van der Waals,
Coulomb and degeneracy effects. In regions where ionization is
complete, non-relativistic or relativistic degeneracy, Coulomb and
exchange effects, as well as Thomas-Fermi contributions at finite temperature
 are taken into account. The liquid and solid phases are also account for in 
the equation of state; however, the release of latent heat is not incorporated
into the evolutionary models.

Neutrino emission rates are taken from \citet{itoh89} with the exception of
the plasma neutrino emission rates which are taken from
\citet{haft94}. Radiative opacities are from OPAL \citep{opal96},
complemented at low temperatures by those of \citet{af94}.  Conductive
opacities are from \citet{hub69,itoh93}. 
More details on LPCODE can be found in \citet{alt03}.

We take the initial abundance to be close to solar, $X_{\rm
init}=0.7055$, $Y_{\rm init}=0.2755$, and $Z_{\rm init}=0.019$.
Convection is treated by means of the mixing length theory and
our choice of the mixing length parameter $\alpha_{\rm MLT}=1.8$ is
representative of the outcome of an extensive calibration of solar
models with varying assumptions on the input physics as well as the
solar parameters carried out by \citet{boot03}.  By default, our
calculations include a mild exponentially decaying diffusive overshoot
at all formal convective boundaries \citep{her97} with the adopted
efficiency of $f=0.016$. Overshooting is suppressed, however, during
the main sequence evolution of models with $M<1.3{\rm M}_\odot$, as
evidence suggests the absence of overshooting in such low-mass stars
when they harbour a small convective core early in their main sequence
(MS) evolution
\citep{sch99,mich04}. Microscopic diffusion is accounted for during the 
hot-white dwarf and white dwarf  phases (see \S~\ref{sec:res} for definition of
the different evolutionary phases considered in this work).

Two complementary mass loss prescriptions are used.  During the
evolution along the Red Giant Branch (RGB) and early Asymptotic Giant
Branch (AGB) the Reimers (1975) law is adopted; while for the
thermally-pulsing AGB (TP-AGB) phase, we use the semi-empirical mass
loss prescription of \citet{mar96}, which relates the mass loss rate
to the pulsational period of the star during the Mira phase and mimics
the superwind phase observed in evolved AGB stars.  The free parameter
$\eta$ in the Reimers prescription is set to 1 in all our calculations
except for the 1~M$_\odot$ case. The precise choice of $\eta$ for
stars more massive than 1.5~M$_\odot$ is, however, of little
consequence because these stars lose most of their mass on the TP-AGB
phase, where the Marigo et al. mass loss prescription is applied. In
the case of the 1~M$_\odot$ model its evolution along the RGB phase is
slow enough that a mass loss rate with $\eta=1$ removes all the
hydrogen envelope before the helium-core is massive enough to ignite,
thus evading the helium-core burning phase and leading to the
formation of a helium-core white dwarf from a single progenitor. To
avoid this non-canonical evolutionary channel, we force a lower mass
loss rate by adopting $\eta=0.3$ for this model. An indirect check for
the mass loss prescription is given by the final white dwarf mass.

We carry out evolutionary calculations from the phase of gravitational
contraction to a pre-main sequence star continuously to a cooled white
dwarf for stars to 6${\rm M}_\odot$. For 7 and 8${\rm M}_\odot$,
however, we skip the latest phase of TP-AGB and the planetary nebula
phase for technical reasons; we resume the
calculation in the late planetary nebula phase.

\section{Results}\label{sec:res}

The physical quantities are presented in Table 1 for 1, 1.5, 1.8, and
2 to 8${\rm M}_\odot$ in one solar mass step, with each case divided
into 7 phases (the 0.8${\rm M}_\odot$ model is given in Table 2 for
the first three phases). A transition from a degenerate helium flash
to gentle helium ignition takes place at around $M\simeq 1.9{\rm
M}_\odot$ and physical features change rather abruptly across this
mass; The two models with 1.5 and 1.8${\rm M}_\odot$, together with
the 1${\rm M}_\odot$ model allow us to study properties of stars that
undergo helium flash. We use $M$ to denote the initial mass of the
star.  Our definition for the 7 phases is:

(A) PMS (pre-main sequence) phase, which is
approximately on the Hayashi track up to the minimum in the stellar
luminosity that follows the reaccomodation of the CN abundances to
their equilibrium values. This point is almost coincident with a local
minimum in the gravitational binding energy of a star.

(B) MS (main sequence), up to the moment when the central mass fraction of
hydrogen drops below 0.001. So, it includes gravitational contraction slightly
after the turn-off from the main sequence.

(C) RGB (red giant branch). It includes subgiant evolution and ends on
the RGB when the stellar luminosity reaches a maximum, coincident with
helium ignition in the stellar core.

(D) He-core. From helium ignition up to the moment when the central 
helium mass fraction drops below 0.001.

(E) AGB (asymptotic giant branch). This phase includes early AGB
(E-AGB) phase and the thermally pulsing phase (TP-AGB).  Post-AGB
evolution including the very first stages of white dwarf evolution (up
to the point where neutrino cooling starts to dominate over photon
radiation) is included in this phase for convenience (this
evolutionary stage is very short and does not contribute significantly
to the overall energetics of stars).

(F) Hot-WD (hot white dwarfs). Evolutionary stage during which energy
loss by neutrino emission is more important than photon energy loss.

(G) WD (white dwarfs), up to the point down to $L/L_\odot=-4.5$ 

The table contains the following 
fundamental evolutionary quantities:
(1) age (in $10^6$ yrs); (2) mass at the end of each step; 
(3) mean luminosity (averaged over the time); (4) mean radius; 
(5) mean effective temperature, and (6) mean surface gravity.  
(2)-(4) are in solar units. 

Integrated energy output (in ergs) from each phase:
(7) $E_\gamma$, energy emitted in photons (integrated bolometric
photon luminosity),
which is broken down into:
(8) $E_{\rm NIR}$, integrated luminosity in the infrared
($\lambda>10000$\AA);
(9) $E_{\rm opt}$, integrated luminosity in the optical band
($3500<\lambda<10000$\AA);
(10) $E_{\rm UV}$, integrated luminosity in the UV
($124<\lambda<3500$\AA);
(11) $E_X$, integrated luminosity in X-rays
($2<\lambda<124$\AA). Note that our numbers are only in thermal radiation
with the assumption of the simple black body spectrum corresponding to
the effective temperature.
For extreme-UV and  X-ray emission, non-thermal and magnetic-field related
processes can be more important in reality; also, reprocessing of photons
by dust is not considered;
(12) $E_{\rm H}$, energy released in photons from hydrogen burning;
(13) $E_{\rm He}$, the same but from helium burning.
For the 8~M$_\odot$ model the nuclear energy release from carbon
burning, $E_{\rm C}$, is listed after $E_{\rm He}$. 
Energies in neutrinos are 
(14) $E_\nu$(nucl), neutrinos emitted in nuclear reactions;
(15) $E_\nu$(therm), pair neutrinos emitted from thermal processes,
such as Compton, plasmon decay, $e^+e^-$ pair annihilation processes etc.

Global energies: 
(16) $\Delta \Omega$, change in gravitational binding energy. The binding energy
is initially adjusted to zero, and $\Delta\Omega$ is added at each
phase; then adding $\Delta\Omega$ over the evolutionary phases gives 
the work required to dissipate stellar material to
infinity; (17) $\Delta U$, change in internal energy. 

Changes in the elemental abundance:
(18) $\Delta M(X)$, net change in hydrogen mass by nuclear reactions;
(19) $\Delta M(Y)$, net change in helium mass by nuclear reactions;
(20) $\Delta M({\rm CNO})$, net change in the CNO elements 
by nuclear reactions; 
(21) $\Delta M({\rm Ne+Mg})$, net change in the Ne and Mg elements 
by nuclear reactions; all these quantities include the mass that is
ejected by mass loss. 
(22)   $\delta M(X)$, hydrogen mass lost by winds;
(23) $\delta M(Y)$,  helium mass lost by winds;
(24) $\delta M({\rm CNO})$, CNO mass lost by winds;
(25) $\delta M({\rm Ne+Mg})$, Ne and Mg mass lost by winds.
Therefore, the helium  mass in a star at the end  of a particular evolutionary
phase, for example, is 
\begin{equation}
M(Y)_{\rm star}=Y_{\rm init}M + \sum\Delta M(Y) -\sum\delta M(Y),
\end{equation}
where the summation extends from PMS up to the particular phase of interest.

Figure 1 shows the evolutionary track for each of our models (as
indicated in each panel). Solid circles A$-$H indicate our
definition of each phase, where A is our initial condition and B is
the zero age main sequence and so on. H is the point at
$L/L_\odot=-4.5$, the terminating point of our calculation.  The table
includes rich information. We summarise in what follows basic features
 from our preliminary reading, although much of the physics involved is 
that can be found in textbooks of stellar evolution.

\subsection{Time-scales}

The evolutionary time-scales are shown in Figure 2 for MS, RGB,
He-core, AGB, Hot-WD and WD phases. For MS and RGB 
the 0.8~M$_\odot$ model is included.  The results from
the Geneva tracks are also plotted for MS and He-core. Good agreement
is seen between Geneva calculations and ours (except for some
difference for the He-core phase of the $M=1.8{\rm M}_\odot$ model).  The
lifetime of MS varies as $t_{\rm MS}\sim M^{-3.5}$ for stars with
fully radiative interiors ($M<1.3{\rm M_\odot}$), changes to $t_{\rm
MS}\sim M^{-3}$ in the range $1.3{\rm M_\odot}<M<2.5{\rm M_\odot}$ and
becomes $t_{\rm MS}\sim M^{-2.3}$ for masses above $3{\rm M_\odot}$.

The variation of lifetime of the He-core phase as a function of mass
shows a glitch between 1.8 and 2${\rm M}_\odot$. The 1.8${\rm
M}_\odot$ model has a degenerate core and thus undergoes a 
helium flash but at flash ignition its helium core is somewhat smaller
(0.432${\rm M}_\odot$) than for less massive stars, 0.462${\rm
M}_\odot$ and 0.468${\rm M}_\odot$ for the 1.5${\rm M}_\odot$ and
1${\rm M}_\odot$ models respectively. The maximum lifetime for the
He-core phase takes place for the 2${\rm M}_\odot$ model, the least
massive model to ignite helium under non-degenerate conditions; 
this model is the least luminous during helium-core burning. 
Across the glitch the lifetime becomes longer by a factor of
2, contrary to the general shortening trend of lifetime as the stellar
mass increases.  It is worth noting that stars just massive enough to
have ignited helium under non-degenerate conditions are expected to
populate a secondary, separated red clump \citep{gir98,gir99}.
The accurate mass range for which such a
distinctive timescale for the He-core phase depends on
details such as overshooting during the main sequence (for example, if
overshooting is not included this mass range
shifts towards higher mass values by $\approx0.3-0.4 {\rm M}_\odot$). 

Good agreement with the Geneva tracks, 
despite many subtle differences in the treatment of
overshooting and in nuclear reaction rates etc., endorses the
reliability of the state-of-the art stellar evolution calculations for
the helium burning phase.

The lifetime of RGB varies as $t_{\rm RGB}\sim M^{-5.5}$ for
$M\le3{\rm M}_\odot$ and $\sim M^{-3.4}$ for higher masses. It is
interesting to note that the lifetimes of AGB and RGB start agreeing
for $>2{\rm M}_\odot$; the two lifetimes almost coincide for $M\ge
4{\rm M}_\odot$, despite the mean luminosity of AGB being higher by a
factor of $3-4$ than that of RGB for this mass range. The lifetime of
AGB is much shorter than that of RGB for $M<2{\rm M}_\odot$.  Early
AGB evolution is much slower than that during the TP-AGB and dominates
the overall AGB time-scales.  Consequently, we expect the uncertainty
in the mass loss rates during AGB evolution to have a negligible
effect on the AGB time-scales.

The lifetime of He-core phase is nearly parallel to that of AGB for
the entire mass range, $1-8{\rm M}_\odot$, approximately one order of
magnitude longer than the latter.

After a short phase of planetary nebulae, stars begin to cool and get
on the white dwarf sequence. The lifetime during the Hot-WD phase,
defined as the time when neutrino cooling dominates over photon
cooling, depends only weakly on the WD mass, ranging from $1\times
10^7$ yrs for the least massive WD model to about $0.8\times 10^7$ yrs
for the most massive one. The integrated neutrino luminosities
are in the range $1.5\pm0.3\times  10 ^{48}$~\hbox{ergs} for the whole WD mass
range. 

The lifetime of WD, which we define by
the time needed from the epoch when photon cooling dominates over 
neutrino cooling to the epoch of
$\log L/L_\odot=-4.5$, is almost constant ($\sim t_{\rm
WD}=7-8\times 10^9$ yrs) for models with $M_{\rm WD}<1{\rm M_\odot}$
but decreases appreciably for more massive models, reaching about
$\sim t_{\rm WD}=4.8\times 10^9$ yrs for the $M_{\rm WD}= 1.15{\rm
M_\odot}$ WD model.  The integrated photon luminosity during the WD
phase ranges between $8-15\times 10^{47}$~\mbox{ergs}. 
Note that contribution from nuclear burning to
the energy budget of the WD decreases monotonically with increasing WD
mass. It is about 22\% for the 1${\rm M}_\odot$ (0.52${\rm M}_\odot$
WD mass), it becomes negligible (about 0.1\%) for the 0.79${\rm M}_\odot$ 
and 0.001\% for the 8${\rm M}_\odot$ (1.15${\rm M}_\odot$ WD mass). 

\subsection{Stellar structure parameters}

The stellar radii may expand by up to a factor of 2 during the
evolution on the main sequence; so in what follows we consider
time-averaged values over the MS. Gravitational binding
energies are also time averages.  The mean radii are plotted in Figure
3 (a) for the main sequence stars as a function of mass. See Table 3
for numbers.

The dependence of stellar radii on the stellar mass is approximated as
\begin{equation}
\log_{10} R= \left\{
\begin{array}{lr}
0.020 + 1.40\log M & {\rm for} \ \ M\le 1.5 \\
0.185 + 0.55\log M & {\rm for} \ \ M\ge 1.5
\end{array} 
\right.
\end{equation}
in solar units.

We define (FP) the structure constant $K$, as
\begin{equation}
\Omega_g=-K\frac{GM^2}{R},
\label{eq:k}
\end{equation}
which is also shown in Figure 3 (b). On the main sequence, the total
average gravitational binding energy varies by a factor of 15 in the
mass range from 1 to 8${\rm M}_\odot$, while $K$ only varies by a factor up to
1.5. $K$ is described approximately by
\begin{equation}
\log_{10} K= \left\{
\begin{array}{lr}
0.21 + 0.8\log M & {\rm for} \ \ M\le 1.5 \\
0.4 - 0.15\log M & {\rm for} \ \ M\ge 1.5. 
\end{array} 
\right.
\end{equation}
We remark that gravitational contraction continues in the MS for stars
with radiative cores ($M\le 1.3{\rm M}_\odot$) while for more massive
stars, i.e. those with a convective core, the modulus of the
gravitational energy slowly decreases with the MS evolution.

The similar quantities averaged over the WD phase are shown in Figure
3 (c) and (d) as a function of the final white dwarf mass. The models
shown correspond to 10 models of initial 1 to 8 ${\rm M}_\odot$. 
The gap between the 0.6${\rm M}_\odot$ white dwarf model
(3${\rm M}_\odot$ progenitor) and the 0.79${\rm M}_\odot$ model
(4${\rm M}_\odot$ progenitor) is commented on in
\S~\ref{sec:massloss}.  Panel (d) shows that the structure factor
$K$ (for which $M$ is replaced with the white dwarf mass $m_f$) is
basically constant, $K=0.95\pm0.02$, with a weakly increasing trend:
the gravitational energy changes by a factor larger than 5
across this mass range.

For other evolutionary phases $K$ varies wildly with the mass, 
meaning that the density
profile of stars are strongly mass-dependent. Table 3 gives the mean 
$R$ and $K$ for the phases of MS, RGB, He-core and WD, 
where $K$ is defined by equation
(\ref{eq:k}) with $M$ replaced by the mass in the relevant phase.

\subsection{Energy output}

Figure 4 shows the photon energy output integrated over the time of
each phase (and the sum) as a function of stellar mass in logarithmic
scales. This shows that it is enough to consider MS, RGB, He-core and
AGB for global energetics. The total energy of radiation increases
 from $E_{\rm photon}=5\times 10^{51}$ for the $1{\rm M}_\odot$ model 
to $20\times 10^{51}$ for $8{\rm M}_\odot$.

The energy output from PMS is approximately 1/160 that from
MS. Approximately 3/4 of its energy is derived from gravitational
contraction. Optical emission is intrinsically more important than IR
(we are ignoring here the dust around those nascent stars). The energy emitted
in the IR band is $30-40$\% of the total energy for $1{\rm M}_\odot$, but
its importance decreases with increasing $M$.

Now we pay some attention to energies produced in the four major
phases.  In three panels of Figure 5 we present the energy output in
different arrangements to stress various aspects of energy production. 
We show the relative importance of energy output in each phase of the
evolution in Panel (a).  At low mass $M<1.5{\rm M}_\odot$ MS gives
only a subdominant contributions, 20$-$30\% the total, and RGB is by
far the important energy source.  The importance of RGB phase,
however, drops sharply by $M=2{\rm M}_\odot$, and MS becomes the
dominant energy source. The energy fraction of RGB drops from 55\% at
1${\rm M}_\odot$ to 5\% at 2${\rm M}_\odot$, and for $M\ge3{\rm
M}_\odot$ it becomes a negligible amount ($<$2\%).  
This is understood by the following consideration: for $M<2{\rm M}_\odot$
models the helium core mass at the end of the main sequence phase is small
and substantial hydrogen shell burning is needed to prepare the helium core
of $m_c\approx 0.47{\rm M}_\odot$, the critical mass for helium
ignition under the degenerate condition. This contrasts to the case for
$M>2{\rm M}_\odot$, for which the star leaves the main sequence with 
a substantial helium core mass, and there is no need to increase
the core mass substantially before helium ignition; the time scale
of RGB is more importantly governed by the Kelvin-Helmholtz contraction
time-scale, during which the temperature rises to $10^8$K. This brings a 
drastic drop of the energy emitted in the RGB phase across the transition 
mass.

He-core burning phase
and AGB each produces 1/2$-$1/3 the amount of energy MS produces. The
relative importance of AGB shows a gradual decline towards higher masses
for $M\ge 2{\rm M}_\odot$.  We note that, while AGB is not
bolometrically important, it becomes dominant over the other phases 
for $M\approx (2-3){\rm M}_\odot$ stars, if one is
concerned with the near infrared light.  The synthetic $K$ band light
over  stellar  populations  is  thus  disturbed strongly  by  AGB  stars  (see
\citealp{bru03}). For lower mass stars the near infrared light is dominated by
RGB and for higher mass by He-core stars.

It would also be useful to see the energy generation from a different
view. Panel (b) shows the energy output per initial mass,
$E_\gamma/M$, i.e., the efficiency as an energy producing engine.  The
overall efficiency drops sharply from 1 to 3${\rm M}_\odot$, and
varies only slowly for $M>3{\rm M}_\odot$ The breakdown to the phases
shows that the efficiency of the MS phase increases, but only
gradually from 1${\rm M}_\odot$ to higher mass.  This represents what
is often characterised as ``when 10\% of hydrogen is consumed, MS
evolves off the main sequence''. In fact, the consumed hydrogen is
9.7\% of the total mass at 1${\rm M}_\odot$. It increases to 17\% at
8${\rm M}_\odot$. We note that this statement is {\it not} valid for
lower mass stars: a kink is observed in $E_\gamma/M$ at 1${\rm
M}_\odot$, which means the efficiency turning to increase to lower
mass. We confirmed in separate MS calculations with a finer mass mesh
that this kink in fact takes place at 1.3 ${\rm M}_\odot$ where the
core changes from radiative to convective; there is actually a small
jump at this mass (not manifest in the figure) above which we assumed
convective overshooting. Even if we do not take overshooting for 1.3
$M\ge {\rm M}_\odot$, the kink still exists at the same mass but
without a jump. For masses smaller than 1 ${\rm M}_\odot$, what is
constant is the integrated energy output $E_\gamma$ and not
$E_\gamma/M$ (see Figure 4).

For stars in which helium flash takes place ($M\le 1.8{\rm M}_\odot$),
the integrated energy output from RGB stays at constant at $3\times
10^{51}$ ergs for $M\lesssim 1{\rm M}_\odot$ and decreases rather
weakly to the maximum mass. This is due to the fact that the
helium-core mass when the flash ignites is almost independent of the
initial mass of the star for $M\lesssim 1{\rm M}_\odot$ and it weakly
decreases as the initial mass increases.  Therefore $E_\gamma/M$
decreases somewhat faster than $1/M$.  It drops rapidly as the
degeneracy is lost in the core.
The energy output is nearly mass independent for high mass stars with
non-degenerate cores: the luminosity and the lifetime nearly
compensate.
 
For the helium burning phase (He-core), $E_\gamma/M$ is nearly
constant for all masses at $0.5\times 10^{51}$ ergs per solar mass. It
gives 20\% of total energy for $M>2{\rm M}_\odot$.

Figure 5(c) shows the fractional energy production from helium burning
and its subcomponent from the He-core phase.  The fractional total energy
production from helium burning decreases gently with the mass: it is
15\% for low mass ($M\le 2 {\rm M}_\odot$) stars and 10\% for high mass
stars. In helium core burning phase of 1${\rm M}_\odot$ stars, helium
burning produces 1.8 time more energy than hydrogen burning. At 2
${\rm M}_\odot$ this factor drops to 0.3 and starts increasing again to
0.5-0.6 for $M>3{\rm M}_\odot$.  In the AGB phase helium burning is the most
important energy source except for $2-3 {\rm M}_\odot$ where H burning
energy is larger. It is worth  noting that the dominant contribution to helium
burning during the AGB phase comes from the E-AGB.

The energetic contribution of the most advanced evolutionary stages,
after the track of AGB stars reaches the maximum luminosity, 
is insignificant ($<0.1$\%),
including that of the long white dwarf phase.

\subsection{Neutrino production}

Figure 6 shows the fraction of energy in neutrinos through the life of
stars. The two open symbols represent neutrinos produced by nuclear
reactions and those of thermal origin. The latter stay insignificant
for the mass range that concerns us; they begin to rise only at 8${\rm
M}_\odot$.  The gradual increase of the neutrino energy fraction from 1
to 3${\rm M}_\odot$ represents an increasing importance of
the later part of the $pp$ chain and then the dominance of the CNO
cycle.  Hydrogen shell burning is always dominated by the CNO
cycle. We expect 6.4\% of energy fractions from beta decay of $^{13}$N
and $^{15}$O for the CN cycle. Neutrinos from NO cycle, originating in
the beta decay of $^{17}$F, have higher energy, but because the
branching between CN and the NO cycles is almost independent of
temperature, they are always a minor contributor.  The nuclear
neutrino energy fraction actually seen in Figure 6 is approximately
10\% less than the value quoted above because $L_\gamma$ includes 
the energy production by helium burning, which produces a negligible
amount of neutrinos  (helium burning is an important  source of neutrinos only
at the helium flash, see Serenelli \& Fukugita 2005). 

The Hot-WD and WD phases give only insignificant contributions 
for neutrinos at several times $10^{48}$ erg. The dominant
neutrino production is from plasmon decay. The integrated neutrino
energy is comparable to that of photons for stars with $M\le3{\rm M}_\odot$. 
For high mass stars energies carried away by neutrinos win but only by
a factor of 2.

\subsection{Mass loss and remnants}  
\label{sec:massloss}

After some nuclear reprocessing, stars return part of their stellar
matter to the interstellar medium.  The mass lost by winds in our
models is shown in Figure 7, where the total, hydrogen, helium and
total metal masses lost per unit initial mass are shown. To help
understanding the amount of matter that underwent nuclear
reprocessing, the equivalent amount that would be expected if
reprocessing of the stellar matter would be switched  off is shown by
dotted curves.  The difference
between the solid and dotted curves is the effect of nuclear
reprocessing in stars.  In our calculation, $>95$\% of mass is
retained up to the beginning of AGB for $M\ge 2{\rm M}_\odot$
stars. For 1${\rm M}_\odot$ stars 25\% of mass is already lost in the
RGB phase and another 25\% (relative to initial mass) in the AGB
phase. More than 80\% of mass is lost by winds for $M>3{\rm M}_\odot$.

The composition of the lost mass depends appreciably on the stellar
mass. The hydrogen and helium abundances are mostly determined by the
increasing efficiency with mass of the first dredge up event for stars
with masses below $\sim 4{\rm M}_\odot$ and by both the first and second
dredge up for masses above this value. While these events alter the
distribution of the CNO elements (nitrogen is enhanced while carbon
and oxygen are depleted), the total metallicity Z of the stellar
envelope is hardly affected. The increase in the metallicity of the
lost mass occurs during the third dredge up, where material partially
processed by helium-shell burning (mostly carbon and to a lesser
extent oxygen) is mixed into the envelope during the TP-AGB phase.
The efficiency of the third dredge up in our models is most noticeable
for stars with masses below $\sim 4{\rm M}_\odot$ as can be seen in Figure
7.  For the range of masses considered in this work, the added
contribution of neon, magnesium and heavier elements to the envelope
metallicity remains almost constant and equal to its initial value.

The remaining mass forms a white dwarf. The initial mass ($M$) - final
mass ($m_f$) relation (IFMR) gives a check for the calculation of the mass
loss.
Figure 8 shows the IFMR, with \citet{wei00}'s semi-empirical relation added for comparison. They agree to
within 5\%, except at the initial mass $3 {\rm M}_\odot$, for which
our calculation gives a white dwarf mass 15\% lower than Weidemann's
value. The ``anchor point'' of Weidemann, $m_f=0.8{\rm M}_\odot$ at
$M=4{\rm M}_\odot$, is met nearly exactly in our calculation, which
gives $m_f=0.79{\rm M}_\odot$.

We may ascribe the particularly small mass, $m_f=0.6{\rm M}_\odot$, at
$M=3{\rm M}_\odot$ to convective  overshooting during the TP-AGB phase
that is  effective at the inner  front of the  helium convective shell
and  pushes   the  helium  shell  inwards  during   a  thermal  pulse,
effectively  preventing  an  increase  in  size of  the  CO-core.   If
overshooting is  suppressed or  if its efficiency  is limited  at this
specific  convective boundary, then  the final  mass for  our $M=3{\rm
M}_\odot$  could  be  about  0.68${\rm  M}_\odot$  in  agreement  with
Weideman's value.\footnote{It should be noted, however, that an amount
of  overshooting  operating  at  the  inner  boundary  of  the  helium
convective shell  similar to  the one adopted  in this paper  (or some
other mechanism  providing extra mixing)  is necessary to  explain the
surface  abundances of  PG-1159  stars \citet{her99}.}   We note  that
theoretical models  used by \citet{wei00}  to derive his  IFMR involve
either synthetic AGB evolution  (e.g. \citealp{gir00}) or full stellar
evolution models  that do  not include overshooting  \citep{blo95} and
this prevents a detailed comparison with our results. While
we do not fully understand  the origin of the discrepancy, 
we believe that the  inclusion  of overshooting  beyond  the  formal  
Schwarzschild
convective  boundaries in  the  helium convective  shell  lies at  its
roots.  The current observational data (e.g., see, Claver et al. 2001)
show too  large a  scatter and we  cannot distinguish between  the two
cases.  This  effect of overshooting becomes weak  already at $M=4{\rm
M}_\odot$  and negligible  towards the  high mass  end covered  by our
models.

In Figure 8 a line  $m_f=0.085M+0.44$, which is close to the empirical
mass relation of  Claver et al.  (2001), is drawn  for a reference; it
describes the  $M-m_f$ relation reasonably well, again  except for the
dip  at $M=3{\rm  M}_\odot$.   At the  high  mass end,  our $M=  8{\rm
M}_\odot$ model leads to a $m_f=1.15{\rm M}_\odot$ WD with a ONe core,
very close to  the minimum mass for core-collapse  that will likely be
about  $9{\rm M}_\odot$.  We  note  that our  white  dwarf masses  are
smaller than  that of  \citet{gir00} for a  given initial  mass. Their
$m_f$  reaches $1.15{\rm  M}_\odot$ for  $M=5{\rm M}_\odot$,  thus the
onset  of core  collapse  supernovae takes  place  at $M\approx  6{\rm
M}_\odot$.  We  note, however,  that  the  synthetic AGB  calculations
employed by \citet{gir00} predict a substantial growth of the hydrogen
depleted  core during TP-AGB  evolution, in  contradiction with
full evolutionary models of TP-AGB stars which, particularly for stellar 
masses $\gtrsim4{\rm  M}_\odot$, predict  a very  small or  even negligible
increase in the core size (i.e. this paper, \citealp{blo00,her00}).

\section{Summary}

The central result of this paper is Table 1, which documents products
of stars integrated over time for each phase. We consider stars from 1
to 8 ${\rm M}_\odot$. This can be used to link baryon mass locked into stars
to extragalactic background light, and also to predict chemical
enrichment.  The table also tells us the relative importance of each phase
for the light emitted by stars.  Some discussion is made to
understand the trend of numbers we present in the table.

When the numbers in the table are integrated over the initial mass
function, they give the output of photon and neutrino energies from
respective phases of stellar evolution and helium and CNO enrichment
per given mass of cooled baryons, although the calculation still must be
supplemented by a similar table for more massive stars. When they are
further integrated over star formation history, we can estimate
differential contributions to the energetics of the galaxies and
Universe. Such procedures will make the work similar to that of 
population sysnthesis model, the most recent example of which is 
Bruzual \& Charlot (2003). The difference lies in the fact that the
latter is specifically oriented for the calculation
of the detailed spectral output of the stellar radiation,
while our aim is to calculate all outputs in a gross form 
 from the stellar population divided into each stage.

\acknowledgments

We would like to thank Jim Peebles for useful discussions from time to
time. AMS  is supported  by  the NSF  (grant  PHY-0503684), the  W. M.  
Keck
Foundation through a grant-in-aid  to the Institute for Advanced Study
and  by the  Association  of  Members of  the  Institute for  Advanced
Study. MF is  supported by the Monell Foundation  at the Institute for
Advanced Study,  and by Grant-in-Aid  of the Ministry of  Education at
the University of Tokyo.

\begin{figure*}
\includegraphics[width=19cm]{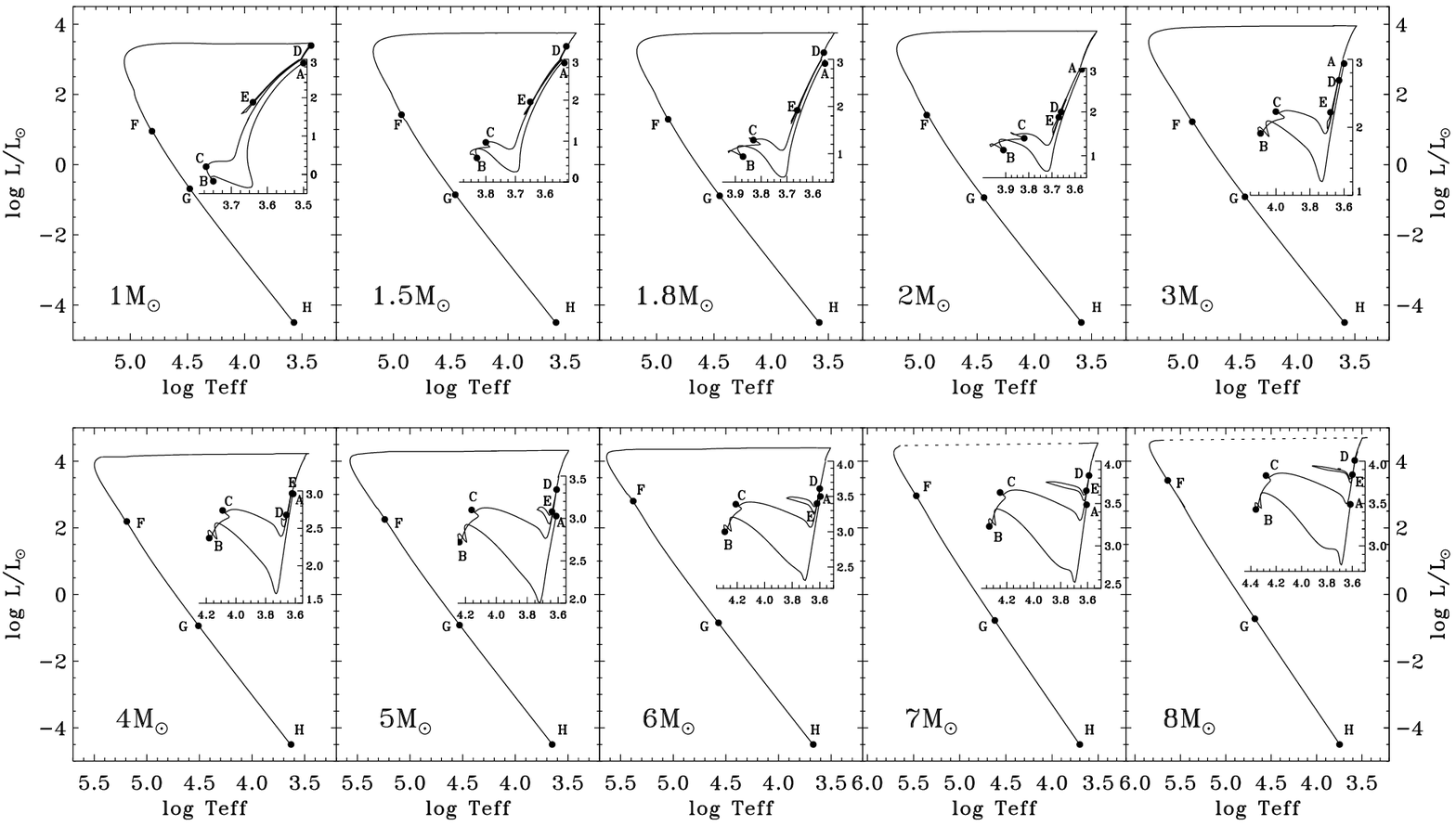}
\caption{Evolutionary tracks for our models. A to H stand for the
points for the six phases defined in the text. In particular, B-C is
the main sequence and D-E is helium core burning. The dotted line
segment for the 7 and $8{\rm M}_\odot$ models is the part we skipped
calculations as described in the text. \label{fig:track}}
\end{figure*}

\begin{figure*}
\includegraphics[width=9cm, bb=40 0 453 311]{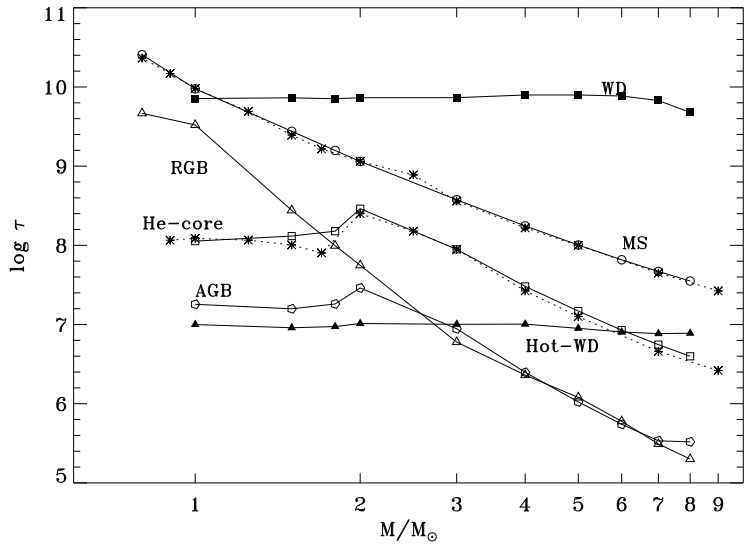}
\caption{Lifetime of each phase (indicated by the legend) as a
function of mass. For the main sequence and the RGB phase the
0.8${\rm M}_\odot$ model is added. Asterisks and dotted lines are
the lifetime from the Geneva track.\label{fig:tscales}}
\end{figure*}

\begin{figure*}
\includegraphics[width=9cm, bb=0 50 396 570]{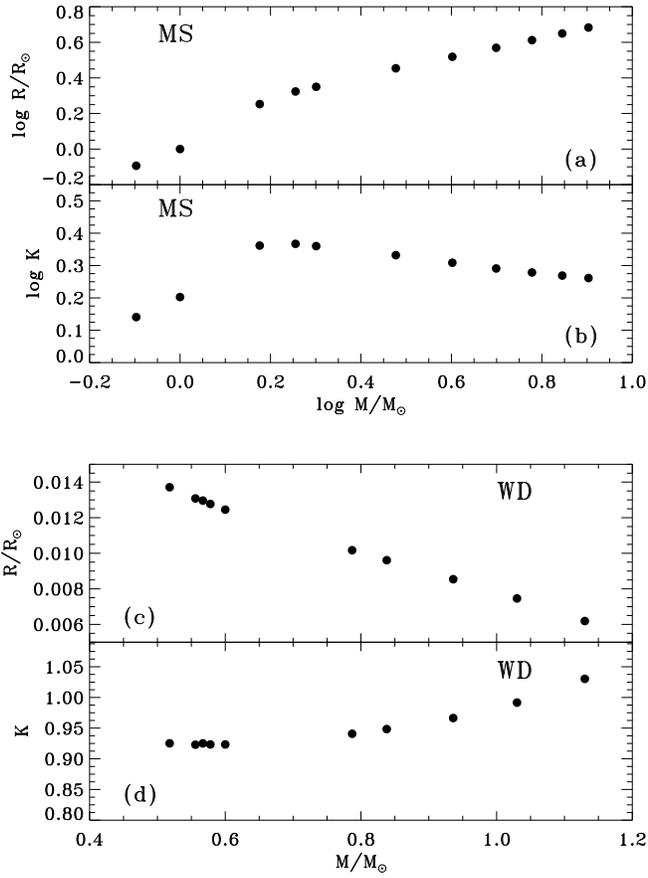}
\caption{Stellar radius and the structure constant as a function of
the mass. Panels (a) and (b) are for main sequence stars (in
logarithmic scales), and (c) and (d) for white dwarfs (in linear
scales).\label{fig:rk}}
\end{figure*}

\begin{figure*}
\includegraphics[width=8cm, bb=40 0 355 270]{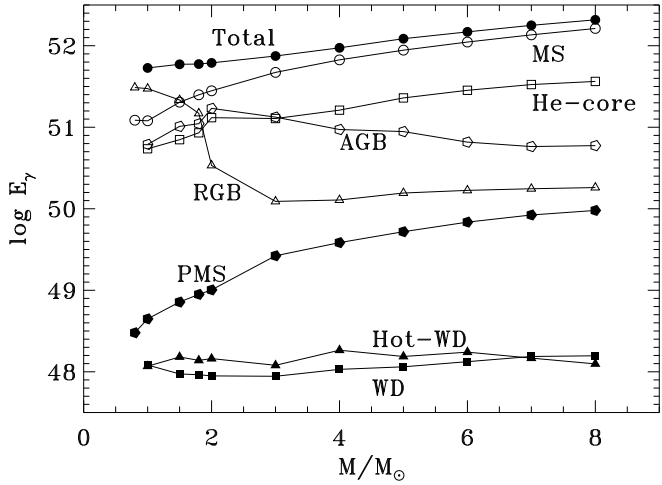}
\caption{Photon energy output in the life of stars and the breakdown
into each phases (indicated by the legend) plotted as a function of stellar mass. $E_\gamma$ is in units of ergs.\label{fig:ephot}}
\end{figure*}

\begin{figure*}
\includegraphics[width=6.5cm, bb=20 20 350 550]{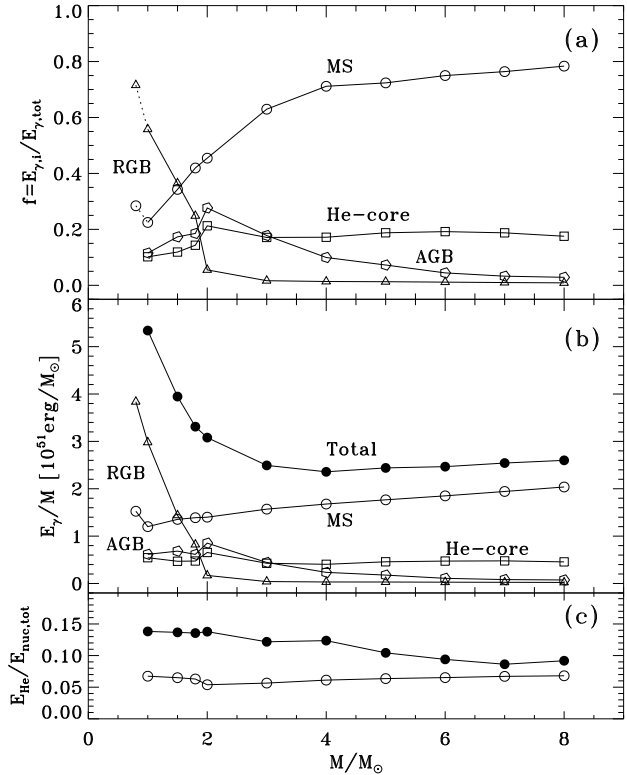}
\caption{(a) Fractional photon energy output from the four major
phases.  The 0.8${\rm M}_\odot$ model is added for the MS and RGB phases.
(b) Photon energy production per initial stellar mass.  (c) Fraction of the
photon energy production from helium burning. The open symbols are the
contribution from the helium core burning phase.\label{fig:fracener}}
\end{figure*}

\begin{figure*}
\includegraphics[width=8cm, bb=20 0 350 275]{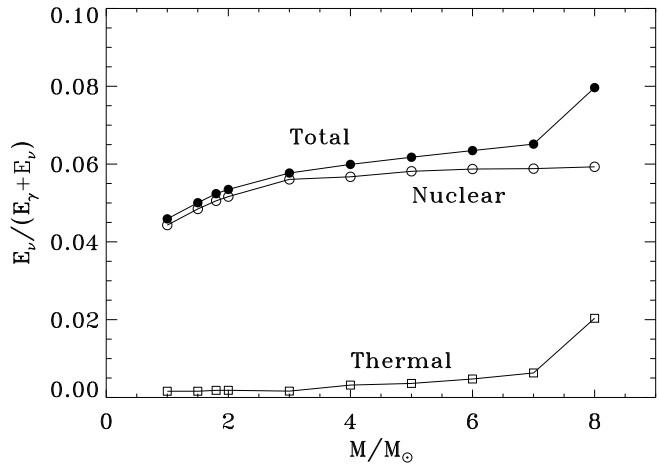}
\caption{Fraction of neutrino energy integrated over the entire life
of stars. The open circles are neutrino energy from nuclear reaction
processes and the open squares are from the thermal pair-neutrino
production.\label{fig:neufrac}}
\end{figure*}

\begin{figure*}
\includegraphics[width=8cm, bb=30 0 350 275]{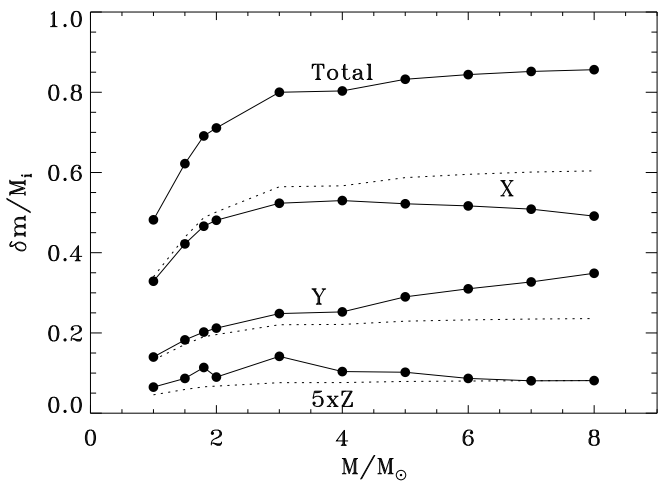}
\caption{Fractional mass loss of stars (relative to the initial
stellar mass). X, Y and Z are breakdown into hydrogen, helium and
heavy elements (scaled up by a factor of 5). The dotted curves are
the equivalent amount when the stellar matter would not be reprocessed by
nuclear reactions.\label{fig:mloss}}
\end{figure*}

\begin{figure*}
\includegraphics[width=8cm, bb=30 0 350 275]{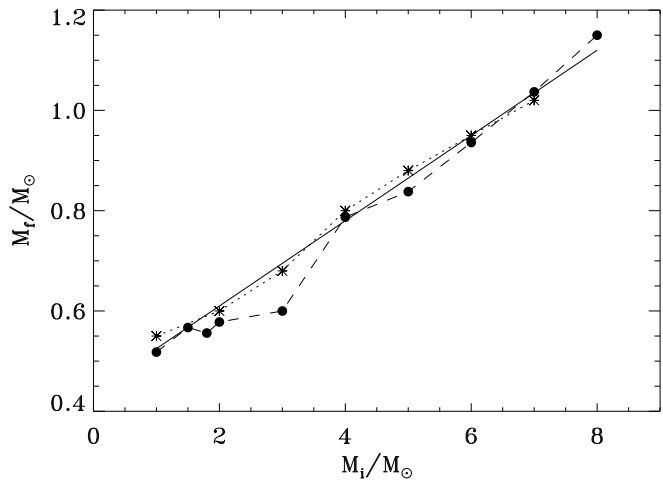}
\caption{Initial-final mass relation. Solid circles stand for our calculation.
Asterisks are the semi-empirical relation of Weidemann. The solid line is
the relation $m_f=0.085M+0.44$.\label{fig:mimf}}
\end{figure*}

\clearpage

\begin{deluxetable}{lrrrrrrrrr}
\tabletypesize{\small}
\tablewidth{0pt}
\tablecolumns{9}
\tablecaption{Evolutionary  characteristics   of  low  and   intermediate  mass
  stars. \label{tab:evol}} 
\tablehead{ \colhead{} & \colhead{PMS} &
\colhead{MS}   &  \colhead{RGB}  &   \colhead{He-core}  &   \colhead{AGB}  &
  \colhead{Hot-WD} & \colhead{WD} & \colhead{Total}} 
\startdata
\multicolumn{9}{l}{$1{\rm M_\odot}$} \\ \hline
Age & 46.5 & 9477 & 12789 & 12902 & 12920 & 12930 & 20074 & --- \\
Mass & 1.0 & 0.995 & 0.751 & 0.744 & 0.518 & 0.518 & 0.518 & --- \\
log $L/L_\odot$ & $-$0.15 & 0.00 & 0.91 & 1.60 & 2.47 & 0.28 & $-$2.85 & --- \\
$R$ & 1.09 & 1.00 & 2.60 & 9.74 & 37.3 & 0.0191 & 0.0137 & --- \\
log $T_{\rm eff}$ & 3.71 & 3.76 & 3.74 & 3.67 & 3.60 & 4.67 & 3.77 & --- \\
log $g_{\rm sup}$ & 4.41 & 4.45 & 4.05 & 2.34 & 1.51 & 7.60 & 7.88 & --- \\
\hline 
$E_\gamma$ & 4.46E+48 & 1.20E+51 & 2.98E+51 & 5.44E+50 & 6.15E+50 & 1.17E+48 & 1.21E+48 & 5.34E+51 \\
\phm{.....} $E_{\rm NIR}$ & 1.59E+48 & 3.37E+50 & 1.55E+51 & 2.25E+50 & 3.65E+50 & 1.30E+45 & 5.38E+46 & 2.47E+51 \\
\phm{.....} $E_{\rm Opt}$ & 2.68E+48 & 7.76E+50 & 1.38E+51 & 3.07E+50 & 2.34E+50 & 2.30E+46 & 2.71E+47 & 2.70E+51 \\
\phm{.....}  $E_{\rm UV}$ & 1.88E+47 & 8.48E+49 & 5.69E+49 & 1.22E+49 & 1.66E+49 & 1.14E+48 & 8.87E+47 & 1.72E+50 \\
\phm{.....} $E_{\rm X}$ & --- & --- & --- & --- & 9.91E+45 & 2.10E+42 & --- & 9.91E+45 \\
$E_{\rm H}$ & 1.47E+48 & 1.20E+51 & 2.97E+51 & 1.96E+50 & 2.12E+50 & 3.29E+47 & 2.65E+47 & 4.59E+51 \\
$E_{\rm He}$ & --- & --- & 1.69E+47 & 3.59E+50 & 3.76E+50 & 2.86E+45 & --- & 7.36E+50 \\
$E_\nu$(nucl) & 4.28E+46 & 3.20E+49 & 1.88E+50 & 1.31E+49 & 1.51E+49 & 2.08E+46 & 7.29E+45 & 2.48E+50\\
$E_\nu$(ther) & --- & 1.42E+43 & 2.62E+47 & 7.13E+47 & 6.48E+48& 1.32E+48 & 3.77E+47 & 9.15E+48 \\
$\Delta \Omega$ & -5.97E+48 & $-$2.35E+47 & $-$3.47E+49 & 2.48E+49 & $-$4.53E+49 & $-$5.03E+48 & $-$3.00E+48 & $-$6.95E+49 \\
$\Delta U$ & 2.98E+48 & 1.24E+47 & 1.82E+49 & $-$1.32E+49 & 2.40E+49 & 2.88E+48 & 1.68E+48 & 3.66E+49 \\
$\Delta M$(X) & $-$1.66E$-$04 & $-$9.66E$-$02 & $-$2.40E$-$01 & $-$1.64E$-$02 & $-$1.80E$-$02& $-$2.91E$-$05 & $-$2.43E$-$05 & $-$3.72E$-$01 \\
$\Delta M$(Y) & 1.57E$-$04 & 9.65E$-$02 & 2.40E$-$01 & $-$2.32E$-$01 & $-$2.25E$-$01& 2.61E$-$05 & 2.23E$-$05 & $-$1.20E$-$01 \\
$\Delta M$(CNO) & 8.67E$-$06 & 1.40E$-$04 & $-$2.54E$-$04 & 2.43E$-$01 & 2.38E$-$01& 3.00E$-$06 & --- & 4.81E$-$01 \\
$\Delta M$(Ne+Mg) & --- & --- & --- & 5.37E$-$03 & 4.73E$-$03 & --- & --- & 1.01E$-$02 \\
$\delta$(X) & 1.56E$-$05 & 3.32E$-$03 & 1.70E$-$01 & 4.89E$-$03 & 1.54E$-$01 & --- & --- & 3.29E$-$01 \\
$\delta$(Y) & 6.07E$-$06 & 1.30E$-$03 & 7.06E$-$02 & 2.12E$-$03 & 6.68E$-$02 & --- & --- & 1.40E$-$01 \\
$\delta$(CNO) & 3.05E$-$07 & 6.51E$-$05 & 3.38E$-$03 & 9.89E$-$05 & 3.13E$-$03 & --- & --- & 6.61E$-$03 \\
$\delta$(Ne+Mg) & 5.19E$-$08 & 1.11E$-$05 & 5.75E$-$04 & 1.68E$-$05 & 5.31E$-$04 & --- & --- & 1.12E$-$03 \\
\hline
\tablebreak 
\multicolumn{9}{l}{$1.5{\rm M_\odot}$} \\ \hline
Age & 15.0 & 2756 & 3033 & 3163.8 & 3179.8 & 3188.9 & 10491 \\
Mass & 1.5 & 1.5 & 1.197 & 1.174 & 0.567 & 0.567 & 0.567 \\
log $L/L_\odot$ & 0.60 & 0.79 & 1.81 & 1.69 & 2.73 & 0.15 & $-$2.97 \\
$R$ & 2.04 & 1.79 & 10.0 & 10.8 & 49.9 & 0.0164 & 0.0130 \\
log $T_{\rm eff}$ & 3.75 & 3.83 & 3.70 & 3.67 & 3.60 & 4.63 & 3.78 \\
log $g_{\rm sup}$ & 4.08 & 4.14 & 3.31 & 2.45 & 1.66 & 7.77 & 7.97 & --- \\
\hline 
$E_\gamma$ & 7.17E+48 & 2.03E+51 & 2.16E+51 & 7.04E+50 & 1.02E+51 & 1.52E+48 & 9.42E+47 & 5.92E+51 \\
\phm{.....}  $E_{\rm NIR}$ & 2.44E+48 & 4.18E+50 & 1.13E+51 & 2.91E+50 & 6.72E+50 & 1.59E+45 & 5.08E+46 & 2.52E+51 \\
\phm{.....}  $E_{\rm Opt}$ & 4.24E+48 & 1.34E+51 & 9.99E+50 & 3.98E+50 & 3.46E+50 & 2.77E+46 & 2.53E+47 & 3.08E+51 \\
\phm{.....}  $E_{\rm UV}$ & 4.94E+47 & 2.75E+50 & 2.68E+49 & 1.56E+49 & 8.43E+48 & 1.49E+48 & 6.39E+47 & 3.34E+50 \\
\phm{.....} $E_{\rm X}$ & --- & --- & --- & --- & 2.81E+46 & 3.46E+43 & --- & 2.81E+46 \\ 
$E_{\rm H}$ & 1.76E+48 & 2.03E+51 & 2.15E+51 & 3.30E+50 & 5.57E+50 & 2.88E+47 & 1.21E+47 & 5.07E+51 \\
$E_{\rm He}$ & --- & --- & 1.75E+47 & 3.82E+50 & 4.21E+50 & 3.26E+45 & --- & 8.03E+50 \\ 
$E_\nu$(nucl) & 7.62E+46 & 9.77E+49 & 1.43E+50 & 2.28E+49 & 3.77E+49 & 1.84E+46 & 3.27E+45 & 3.02E+50 \\
$E_\nu$(ther) & --- & 2.77E+43 & 2.72E+47 & 8.82E+47 & 7.53E+48 & 2.09E+48 & 1.85E+47 & 1.10E+49 \\
$\Delta \Omega$ & $-$1.08E+49 & 6.72E+47 & $-$2.93E+49 & 2.25E+49 & $-$5.99E+49 & $-$7.94E+48 & $-$2.36E+48 & $-$8.71E+49 \\
$\Delta U$ & 5.39E+48 & $-$3.05E+47 & 1.48E+49 & $-$1.15E+49 & 3.08E+49 & 4.63E+48 & 1.36E+48 & 4.52E+49 \\
$\Delta M$(X) & $-$1.81E$-$04 & $-$1.67E$-$01 & $-$1.78E$-$01 & $-$2.63E$-$02 & $-$4.64E$-$02 & $-$2.54E$-$05 & $-$1.13E$-$05 & $-$4.18E$-$01 \\
$\Delta M$(Y) & 1.67E$-$04 & 1.67E$-$01 & 1.78E$-$01 & $-$2.33E$-$01 & $-$2.40E$-$01 & 2.48E$-$06 & 1.25E$-$05 & $-$1.27E$-$01 \\
$\Delta M$(CNO) & 1.30E$-$04 & 1.87E$-$05 & $-$1.12E$-$04 & 2.53E$-$01 & 2.80E$-$01 & 6.74E$-$07 & --- & 5.33E$-$01 \\
$\Delta M$(Ne+Mg) & --- & --- & --- & 5.62E$-$03 & 5.62E$-$03 & --- & --- & 1.12E$-$02 \\
$\delta$(X) & --- & --- & 2.08E$-$01 & 1.60E$-$02 & 4.16E$-$01 & --- & --- & 6.40E$-$01 \\
$\delta$(Y) & --- & --- & 8.88E$-$02 & 6.86E$-$03 & 1.78E$-$01 & --- & --- & 2.74E$-$01 \\
$\delta$(CNO) & --- & --- & 4.22E$-$03 & 3.25E$-$04 & 8.46E$-$03 & --- & --- & 1.30E$-$02 \\
$\delta$(Ne+Mg) & --- & --- & 7.12E$-$04 & 5.48E$-$05 & 1.43E$-$03 & --- & --- & 2.20E$-$03 \\ 
\hline
\tablebreak 
\multicolumn{9}{l}{$1.8{\rm M_\odot}$} \\ \hline
Age & 9.54 & 1579 & 1678 & 1829 & 1847.4 & 1856.6 & 8964 \\
Mass & 1.8 & 1.8 & 1.665 & 1.647 & 0.556 & 0.556 & 0.556 \\
log $L/L_\odot$ & 0.92 & 1.12 & 2.09 & 1.69 & 2.69 & 0.10 & $-$2.97 \\
$R$ & 2.38 & 2.11 & 15.6 & 10.1 & 42.7 & 0.0165 & 0.0131 \\
log $T_{\rm eff}$ & 3.81 & 3.89 & 3.69 & 3.68 & 3.62 & 4.62 & 3.78 \\
log $g_{\rm sup}$ & 4.04 & 4.11 & 3.08 & 2.65 & 1.90 & 7.75 & 7.95 & --- \\
\hline 
$E_\gamma$ & 8.90E+48 & 2.50E+51 & 1.48E+51 & 8.54E+50 & 1.10E+51 & 1.38E+48 & 9.16E+47 & 5.96E+51 \\
\phm{.....} $E_{\rm NIR}$ & 3.25E+48 & 4.09E+50 & 7.39E+50 & 3.35E+50 & 6.84E+50 & 1.60E+45 & 5.00E+46 & 2.17E+51 \\
\phm{.....} $E_{\rm Opt}$ & 4.91E+48 & 1.59E+51 & 7.21E+50 & 4.95E+50 & 4.12E+50 & 2.76E+46 & 2.48E+47 & 3.23E+51 \\
\phm{.....} $E_{\rm UV}$ & 7.40E+47 & 4.99E+50 & 1.98E+49 & 2.33E+49 & 1.01E+49 & 1.35E+48 & 6.19E+47 & 5.57E+50 \\
\phm{.....} $E_{\rm X}$ & --- & --- & --- & --- & 2.42E+46 & 1.58E+43 & --- & 2.42E+46 \\ 
$E_{\rm H}$ & 1.98E+48 & 2.51E+51 & 1.47E+51 & 4.91E+50 & 6.49E+50 & 2.59E+47 & 1.14E+47 & 5.12E+51 \\
$E_{\rm He}$ & --- & --- & 1.40E+47 & 3.73E+50 & 4.30E+50 & 2.02E+45 & --- & 8.03E+50 \\
$E_\nu$(nucl) & 9.76E+46 & 1.43E+50 & 9.83E+49 & 3.28E+49 & 4.31E+49 & 1.66E+46 & 3.08E+45 & 3.18E+50 \\
$E_\nu$(ther) & --- & 3.11E+43 & 3.11E+47 & 8.92E+47 & 7.55E+48 & 1.87E+48 & 1.89E+47 & 1.08E+49 \\
$\Delta \Omega$ & $-$1.39E+49 & 1.35E+48 & $-$2.01E+49 & 1.63E+49 & $-$5.72E+49 & $-$7.07E+48 & $-$2.30E+48 & $-$8.30E+49 \\
$\Delta U$ & 6.93E+48 & $-$6.36E+47 & 1.02E+49 & $-$6.33E+48 & 2.97E+49 & 4.09E+48 & 1.31E+48 & 4.16E+49 \\
$\Delta M$(X) & $-$2.02E$-$04 & $-$2.10E$-$01 & $-$1.22E$-$01 & $-$3.76E$-$02 & $-$5.56E$-$02 & $-$2.27E$-$05 & $-$1.08E$-$05 & $-$4.25E$-$01 \\
$\Delta M$(Y) & 1.33E$-$04 & 2.10E$-$01 & 1.22E$-$01 & $-$2.15E$-$01 & $-$2.34E$-$01 & 2.25E$-$05 & 1.08E$-$05 & $-$1.16E$-$01 \\
$\Delta M$(CNO) & 1.10E$-$04 & $-$7.94E$-$05 & $-$5.29E$-$05 & 2.47E$-$01 & 2.83E$-$01 & 1.35E$-$06 & --- & 5.30E$-$01\\
$\Delta M$(Ne+Mg) & --- & --- & --- & 5.33E$-$03 & 6.01E$-$03 & --- & --- & 1.13E$-$02 \\
$\delta$(X) & --- & --- & 9.27E$-$02 & 1.25E$-$02 & 7.41E$-$01 & --- & --- & 8.46E$-$01 \\
$\delta$(Y) & --- & --- & 3.95E$-$02 & 5.40E$-$03 & 3.22E$-$01 & --- & --- & 3.67E$-$01 \\
$\delta$(CNO) & --- & --- & 1.87E$-$03 & 2.54E$-$04 & 2.10E$-$02 & --- & --- & 2.31E$-$02 \\
$\delta$(Ne+Mg) & --- & --- & 3.17E$-$04 & 4.30E$-$05 & 2.75E$-$03 & --- & --- & 3.11E$-$03 \\ 
\hline
\tablebreak 
\multicolumn{9}{l}{$2{\rm M_\odot}$} \\ \hline
Age & 7.5 & 1147 & 1203 & 1493 & 1522 & 1532.3 & 8853 \\
Mass & 2.0 & 1.99 & 1.98 & 1.96 & 0.578 & 0.578 & 0.578 \\
log $L/L_\odot$ & 1.02 & 1.31 & 1.70 & 1.57 & 2.68 & 0.07 & $-$3.00 \\
$R$ & 2.84 & 2.24 & 8.16 & 8.27 & 38.8 & 0.0157 & 0.0128 \\
log $T_{\rm eff}$ & 3.78 & 3.92 & 3.76 & 3.69 & 3.63 & 4.44 & 3.59 \\
log $g_{\rm sup}$ & 3.89 & 4.11 & 3.33 & 2.90 & 2.10 & 7.81 & 7.99 & --- \\
\hline 
$E_\gamma$ & 1.01E+49 & 2.80E+51 & 3.40E+50 & 1.31E+51 & 1.70E+51 & 1.45E+48 & 8.91E+47 & 6.16E+51 \\
\phm{.....} $E_{\rm NIR}$ & 2.39E+48 & 3.87E+50 & 1.25E+50 & 4.88E+50 & 1.03E+51 & 1.59E+45 & 4.98E+46 & 2.03E+51 \\
\phm{.....} $E_{\rm Opt}$ & 6.36E+48 & 1.72E+51 & 1.98E+50 & 7.81E+50 & 6.55E+50 & 2.74E+46 & 2.47E+47 & 3.36E+51 \\ 
\phm{.....} $E_{\rm UV}$ & 1.36E+48 & 6.94E+50 & 1.66E+49 & 4.20E+49 & 1.42E+49 & 1.42E+48 & 5.94E+47 & 7.70E+50 \\
\phm{.....} $E_{\rm X}$ & --- & --- & --- & --- & 3.52E+46 & 4.52E+43 & --- & 3.52E+46 \\
$E_{\rm H}$ & 2.12E+48 & 2.81E+51 & 3.40E+50 & 9.82E+50 & 1.15E+51 & 2.25E+47 & 9.08E+46 & 5.28E+51 \\
$E_{\rm He}$ & --- & --- & 2.27E+47 & 3.30E+50 & 5.13E+50 & 1.44E+45 & --- & 8.44E+50 \\
$E_\nu$(nucl) & 1.26E+47 & 1.71E+50 & 2.29E+49 & 6.56E+49 & 7.68E+49 & 1.44E+46 & 2.45E+45 & 3.36E+50 \\
$E_\nu$(ther) & --- & 3.18E+43 & 2.50E+46 & 9.67E+47 & 8.64E+48 & 2.14E+48 & 1.61E+47 & 1.19E+49 \\
$\Delta \Omega$ & $-$1.60E+49 & 3.36E+48 & $-$9.31E+47 & $-$2.24E+48 & $-$6.60E+49 & $-$8.09E+48 & $-$2.27E+48 & $-$9.21E+49 \\
$\Delta U$ & 8.00E+48 & $-$1.64E+48 & 4.96E+47 & 1.05E+48 & 3.39E+49 & 4.73E+48 & 1.32E+48 & 4.79E+49 \\
$\Delta M$(X) & $-$2.14E$-$04 & $-$2.34E$-$01 & $-$2.95E$-$02 & $-$8.45E$-$02 & $-$1.06E$-$01 & $-$2.03E$-$05 & $-$8.44E$-$06 & $-$4.55E$-$01 \\
$\Delta M$(Y) & 6.00E$-$05 & 2.34E$-$01 & 2.93E$-$02 & $-$1.37E$-$01 & $-$2.40E$-$01 & 2.09E$-$05 & 8.44E$-$06 & $-$1.13E$-$01 \\
$\Delta M$(CNO) & 1.54E$-$04 & $-$1.44E$-$04 & 2.03E$-$04 & 2.17E$-$01 & 3.37E$-$01 & $-$1.03E$-$06 & --- & 5.54E$-$01 \\
$\Delta M$(Ne+Mg) & --- & --- & --- & 4.80E$-$03 & 7.27E$-$03 & --- & --- & 1.21E$-$02 \\
$\delta$(X) & 2.82E$-$05 & 7.46E$-$03 & 5.13E$-$03 & 1.23E$-$02 & 9.37E$-$01  & --- & --- & 9.62E$-$01 \\
$\delta$(Y) & 1.10E$-$05 & 2.91E$-$03 & 2.21E$-$03 & 5.39E$-$03 & 4.14E$-$01  & --- & --- & 4.24E$-$01 \\
$\delta$(CNO) & 5.53E$-$07 & 1.46E$-$04 & 1.04E$-$04 & 2.51E$-$04 & 2.66E$-$02 & --- & --- & 2.71E$-$02 \\
$\delta$(Ne+Mg) & 9.41E$-$08 & 2.49E$-$05 & 1.76E$-$05 & 4.25E$-$05 & 3.50E$-$03 & --- & --- & 3.58E$-$03 \\
\hline
\tablebreak 
\multicolumn{9}{l}{$3{\rm M_\odot}$} \\ \hline
Age & 3.8 & 376 & 382 & 471 & 479.9 & 489.9 & 7816 \\
Mass & 3.0 & 2.99 & 2.98 & 2.96 & 0.600 & 0.600 & 0.600 \\
log $L/L_\odot$ & 1.76 & 2.02 & 2.26 & 2.07 & 3.11 & $-$0.01 & $-$3.00 \\
$R$ & 4.37 & 2.84 & 14.5 & 14.7 & 65.9 & 0.0150 & 0.0124 \\
log $T_{\rm eff}$ & 3.91 & 4.05 & 3.80 & 3.69 & 3.63 & 4.61 & 3.79 \\
log $g_{\rm sup}$ & 3.90 & 4.09 & 3.04 & 2.60 & 1.94 & 7.87 & 8.03 & --- \\
\hline 
$E_\gamma$ & 2.65E+49 & 4.71E+51 & 1.23E+50 & 1.28E+51 & 1.33E+51 & 1.20E+48 & 8.78E+47 & 7.48E+51 \\
\phm{.....} $E_{\rm NIR}$ & 3.70E+48 & 3.33E+50 & 3.82E+49 & 4.81E+50 & 8.26E+50 & 1.39E+45 & 4.88E+46 & 1.68E+51 \\
\phm{.....} $E_{\rm Opt}$ & 1.36E+49 & 2.27E+51 & 7.15E+49 & 7.58E+50 & 5.01E+50 & 2.39E+46 & 2.46E+47 & 3.61E+51 \\
\phm{.....} $E_{\rm UV}$ & 9.13E+48 & 2.12E+51 & 1.35E+49 & 4.01E+49 & 8.20E+48 & 1.17E+48 & 5.84E+47 & 2.19E+51 \\
\phm{.....} $E_{\rm X}$ & --- & --- & --- & --- & 3.70E+46 & 4.14E+43 & --- & 3.70E+46 \\
$E_{\rm H}$ & 1.04E+49 & 4.72E+51 & 1.29E+50 & 8.57E+50 & 8.18E+50 & 1.00E+47 & 5.28E+46 & 6.54E+51 \\
$E_{\rm He}$ & --- & --- & 1.50E+47 & 4.21E+50 & 4.86E+50 & 5.51E+44 & --- & 9.07E+50 \\
$E_\nu$(nucl) & 6.72E+47 & 3.23E+50 & 8.78E+48 & 5.76E+49 & 5.42E+49 & 6.40E+45 & 1.43E+45 & 4.45E+50 \\
$E_\nu$(ther) & --- & 3.84E+43 & 9.61E+44 & 6.52E+47 & 1.02E+49 & 1.91E+48 & 1.42E+47 & 1.29E+49 \\
$\Delta \Omega$ & $-$3.22E+49 & 8.12E+48 & 1.10E+49 & $-$8.17E+48 & $-$7.21E+49 & $-$7.34E+48 & $-$2.34E+48 & $-$1.03E+50 \\
$\Delta U$ & 1.61E+49 & $-$3.99E+48 & $-$5.48E+48 & 4.02E+48 & 3.69E+49 & 4.34E+48 & 1.38E+48 & 5.33E+49 \\
$\Delta M$(X) & $-$9.65E$-$04 & $-$3.97E$-$01 & $-$1.05E$-$02 & $-$7.31E$-$02 & $-$6.85E$-$02 & $-$9.20E$-$06 & $-$4.32E$-$06 & $-$5.50E$-$01 \\
$\Delta M$(Y) & 5.25E$-$04 & 3.97E$-$01 & 1.04E$-$02 & $-$2.16E$-$01 & $-$2.61E$-$01 &  9.20E$-$06 & 4.32E$-$06 & $-$6.86E$-$02 \\
$\Delta M$(CNO) & 4.58E$-$04 & $-$6.34E$-$04 & 1.40E$-$04 & 2.82E$-$01 & 3.21E$-$01 & 3.54E$-$07 & --- & 6.03E$-$01 \\
$\Delta M$(Ne+Mg) & --- & --- & --- & 6.24E$-$03 & 8.18E$-$03 & --- & --- & 1.44E$-$02 \\
$\delta$(X) & 8.47E$-$05 & 1.08E$-$02 & 1.68E$-$03 & 1.50E$-$02 & 1.54E+00 & --- & --- & 1.57E+00 \\
$\delta$(Y) & 3.31E$-$05 & 4.20E$-$03 & 7.11E$-$04 & 6.82E$-$03 & 7.34E$-$01 & --- & --- & 7.45E$-$01 \\
$\delta$(CNO) & 1.66E$-$06 & 2.11E$-$04 & 3.38E$-$05 & 3.08E$-$04 & 7.15E$-$02 & --- & --- & 7.20E$-$02 \\
$\delta$(Ne+Mg) & 2.82E$-$07 & 3.59E$-$05 & 5.73E$-$06 & 5.22E$-$05 & 7.68E$-$03 & --- & --- & 7.77E$-$03 \\
\hline
\tablebreak 
\multicolumn{9}{l}{$4{\rm M_\odot}$} \\ \hline
Age & 1.85 & 176 & 178.3 & 208.6 & 211.09 & 221.2 & 8150 \\
Mass & 4.0 & 3.98 & 3.98 & 3.93 & 0.787 & 0.787 & 0.787 \\
log $L/L_\odot$ & 2.23 & 2.50 & 2.74 & 2.64 & 3.49 & 0.18 & $-$2.96 \\
$R$ & 6.37 & 3.30 & 22.2 & 30.6 & 117 & 0.0113 & 0.0102 \\
log $T_{\rm eff}$ & 3.97 & 4.13 & 3.87 & 3.68 & 3.61 & 4.64 & 3.84 \\
log $g_{\rm sup}$ & 3.85 & 4.08 & 3.03 & 2.11 & 1.47 & 8.23 & 8.32 & --- \\
\hline 
$E_\gamma$ & 3.84E+49 & 6.71E+51 & 1.28E+50 & 1.62E+51 & 9.38E+50 & 1.84E+48 & 1.07E+48 & 9.43E+51 \\
\phm{.....} $E_{\rm NIR}$ & 4.13E+48 & 2.87E+50 & 3.35E+49 & 6.49E+50 & 5.87E+50 & 8.60E+44 & 4.34E+46 & 1.56E+51 \\
\phm{.....} $E_{\rm Opt}$ & 1.67E+49 & 2.47E+51 & 6.91E+49 & 9.26E+50 & 3.47E+50 & 1.51E+46 & 2.52E+47 & 3.83E+51 \\
\phm{.....} $E_{\rm UV}$ & 1.76E+49 & 3.95E+51 & 2.53E+49 & 4.13E+49 & 4.22E+48 & 1.80E+48 & 7.69E+47 & 4.04E+51 \\
\phm{.....} $E_{\rm X}$ & --- & --- & --- & --- & 3.81E+46 & 2.01E+46 & --- & 5.82E+46 \\
$E_{\rm H}$ & 1.51E+49 & 6.72E+51 & 1.37E+50 & 1.04E+51 & 3.04E+50 & 6.34E+46 & 1.76E+46 & 8.22E+51 \\ 
$E_{\rm He}$ & --- & --- & 5.19E+47 & 5.74E+50 & 5.89E+50 & 3.72E+45 & --- & 1.16E+51 \\
$E_\nu$(nucl) & 1.04E+48 & 4.68E+50 & 9.34E+48 & 7.02E+49 & 2.04E+49 & 4.25E+45 & 4.75+44 & 5.69E+50 \\
$E_\nu$(ther) & --- & 4.36E+43 & 5.99E+44 & 5.87E+47 & 2.66E+49 & 4.79E+48 & 1.16E+47 & 3.22E+49 \\
$\Delta \Omega$ & $-$4.65E+49 & 1.11E+49 & 1.82E+49 & $-$1.41E+49 & $-$1.69E+50 & $-$1.89E+49 & $-$3.56E+48 & $-$2.22E+50 \\
$\Delta U$ & 2.33E+49 & $-$5.35E+48 & $-$9.16E+48 & 7.18E+48 & 9.42E+49 & 1.23E+49 & 2.40E+48 & 1.25E+50 \\
$\Delta M$(X) & $-$1.28E$-$03 & $-$5.66E$-$01 & $-$1.18E$-$02 & $-$8.81E$-$02 & $-$2.50E$-$02 & $-$5.71E$-$06 & $-$2.16E$-$06 & $-$6.92E$-$01 \\
$\Delta M$(Y) & 6.72E$-$04 & 5.67E$-$01 & 1.12E$-$02 & $-$3.03E$-$01 & $-$3.69E$-$01 & 4.93E$-$06 & 2.02E$-$06 & $-$9.31E$-$02 \\
$\Delta M$(CNO) & 6.43E$-$04 & $-$9.92E$-$04 & 4.72E$-$04 & 3.83E$-$01 & 3.91E$-$01 & 1.65E$-$06 & --- & 7.74E$-$01 \\
$\Delta M$(Ne+Mg) & --- & --- & --- & 8.28E$-$03 & 8.75E$-$03 & --- & --- & 1.70E$-$02 \\
$\delta$(X) & 1.20E$-$04 & 1.34E$-$02 & 1.85E$-$03 & 2.99E$-$02 & 2.08E+00 & --- & --- & 2.12E+00 \\
$\delta$(Y) & 4.68E$-$05 & 5.24E$-$03 & 7.62E$-$04 & 1.34E$-$02 & 9.86E$-$01 & --- & --- & 1.01E+00 \\
$\delta$(CNO) & 2.35E$-$06 & 2.63E$-$04 & 3.69E$-$05 & 6.11E$-$04 & 6.52E$-$02 & --- & --- & 6.61E$-$02 \\
$\delta$(Ne+Mg) & 4.00E$-$07 & 4.48E$-$05 & 6.25E$-$06 & 1.04E$-$04 & 8.45E$-$03 & --- & --- & 8.60E$-$03 \\
\hline
\tablebreak 
\multicolumn{9}{l}{$5{\rm M_\odot}$} \\ \hline
Age & 1.04 & 101 & 102.2 & 117 & 118.055 &  127.0 & 8048 \\
Mass & 5.00 & 4.98 & 4.97 & 4.89 & 0.838 & 0.838 & 0.838 \\
log $L/L_\odot$ & 2.60 & 2.86 & 3.12 & 3.11 & 3.82 & $-$0.43 & $-$2.92 \\
$R$ & 7.92 & 3.71 & 33.8 & 52.5 & 196 & 0.0104 & 0.00961 \\
log $T_{\rm eff}$ & 4.03 & 4.20 & 3.92 & 3.68 & 3.58 & 4.62 & 3.86 \\
log $g_{\rm sup}$ & 3.86 & 4.07 & 3.04 & 1.75 & 1.02 & 8.33 & 8.40 & --- \\
\hline 
$E_\gamma$ & 5.23E+49 & 8.83E+51 & 1.56E+50 & 2.29E+51 & 8.85E+50 & 1.54E+48 & 1.15E+48 & 1.22E+52 \\
\phm{.....} $E_{\rm NIR}$ & 4.10E+48 & 2.57E+50 & 3.94E+49 & 9.10E+50 & 5.68E+50 & 7.08E+44 & 4.13E+46 & 1.78E+51 \\
\phm{.....} $E_{\rm Opt}$ & 1.94E+49 & 2.56E+51 & 7.61E+49 & 1.31E+51 & 3.14E+50 & 1.24E+46 & 2.52E+47 & 4.28E+51 \\
\phm{.....} $E_{\rm UV}$ & 2.88E+49 & 6.02E+51 & 4.00E+49 & 6.78E+49 & 2.99E+48 & 1.50E+48 & 8.54E+47 & 6.16E+51 \\
\phm{.....} $E_{\rm X}$ & --- & --- & --- & --- & 1.54E+46 & 2.44E+46 & --- & 3.98E+46 \\
$E_{\rm H}$ & 2.13E+49 & 8.85E+51 & 1.66E+50 & 1.51E+51 & 3.45E+50 & 6.44E+45 & 1.64E+45 & 1.09E+52 \\
$E_{\rm He}$ & --- & --- & 1.69E+48 & 7.74E+50 & 4.93E+50 & 3.44E+45 & --- & 1.27E+51 \\
$E_\nu$(nucl) & 1.48E+48 & 6.17E+50 & 1.14E+49 & 1.03E+50 & 2.31E+49 & 4.32E+44 & 4.43E+43 & 7.56E+50 \\
$E_\nu$(ther) & --- & 4.92E+43 & 8.72E+44 & 6.00E+47 & 4.20E+49 & 4.40E+48 & 1.21E+47 & 4.71E+49 \\
$\Delta \Omega$ & $-$6.20E+49 & 1.48E+49 & 2.45E+49 & $-$2.16E+49 & $-$2.03E+50 & $-$1.84E+49 & $-$4.11E+48 & $-$2.70E+50 \\
$\Delta U$ & 3.10E+49 & $-$7.01E+48 & $-$1.25E+49 & 1.12E+49 & 1.16E+50 & 1.25E+49 & 2.86E+48 & 1.55E+50 \\
$\Delta M$(X) & $-$1.89E$-$03 & $-$7.41E$-$01 & $-$1.25E$-$02 & $-$1.28E$-$01 & $-$2.80E$-$02 & $-$5.36E$-$07 & $-$1.23E$-$06 & $-$9.11E$-$01 \\
$\Delta M$(Y) & 1.10E$-$03 & 7.42E$-$01 & 1.13E$-$02 & $-$3.82E$-$01 & $-$3.07E$-$01 & $-$1.14E$-$06 & 1.23E$-$07 & 6.60E$-$02 \\
$\Delta M$(CNO) & 8.34E$-$04 & $-$1.29E$-$03 & 1.22E$-$03 & 4.99E$-$01 & 3.29E$-$01 & 1.76E$-$06 & --- & 8.28E$-$01 \\
$\Delta M$(Ne+Mg) & --- & --- & --- & 1.08E$-$02 & 7.72E$-$03 & --- & --- & 1.85E$-$02 \\
$\delta$(X) & 1.27E$-$04 & 1.60E$-$02 & 2.75E$-$03 & 5.65E$-$02 & 2.53E+00  & --- & --- & 2.61E+00 \\
$\delta$(Y) & 4.96E$-$05 & 6.25E$-$03 & 1.14E$-$03 & 2.50E$-$02 & 1.41E+00  & --- & --- & 1.45E+00 \\
$\delta$(CNO) & 2.49E$-$06 & 3.14E$-$04 & 5.50E$-$05 & 1.15E$-$03 & 8.30E$-$02  & --- & --- & 8.45E$-$02 \\
$\delta$(Ne+Mg) & 4.24E$-$07 & 5.34E$-$05 & 9.33E$-$06 & 1.95E$-$04 & 1.09E$-$02 & --- & --- & 1.12E$-$02 \\
\hline
\tablebreak 
\multicolumn{9}{l}{$6{\rm M_\odot}$} \\ \hline
Age & 0.69 & 65.7 & 66.3 & 74.8 & 75.35 & 83.4 & 7780 \\
Mass & 6.00 & 5.98 & 5.97 & 5.86 & 0.936 & 0.936 & 0.936 \\
log $L/L_\odot$ & 2.91 & 3.15 & 3.42 & 3.44 & 3.98 & 0.25 & $-$2.85 \\
$R$ & 10.2 & 4.09 & 48.6 & 71.8 & 238 & 0.00917 & 0.00854 \\
log $T_{\rm eff}$ & 4.09 & 4.25 & 3.96 & 3.72 & 3.57 & 4.68 & 3.89 \\
log $g_{\rm sup}$ & 3.89 & 4.06 & 3.05 & 1.69 & 0.77 & 8.49 & 8.55 & --- \\
\hline 
$E_\gamma$ & 6.86E+49 & 1.11E+52 & 1.68E+50 & 2.84E+51 & 6.57E+50 & 1.74E+48 & 1.33E+48 & 1.48E+52 \\ 
\phm{.....} $E_{\rm NIR}$ & 4.89E+48 & 2.38E+50 & 4.27E+49 & 1.02E+51 & 4.10E+50 & 4.97E+44 & 3.64E+46 & 1.71E+51 \\
\phm{.....} $E_{\rm Opt}$ & 2.20E+49 & 2.62E+51 & 7.55E+49 & 1.65E+51 & 2.45E+50 & 8.88E+45 & 2.47E+47 & 4.60E+51 \\
\phm{.....} $E_{\rm UV}$ & 4.17E+49 & 8.25E+51 & 4.99E+49 & 1.72E+50 & 2.31E+48 & 1.64E+48 & 1.05E+48 & 8.52E+51 \\
\phm{.....} $E_{\rm X}$ & --- & --- & --- & --- & 1.21E+46 & 9.25E+46 & --- & 1.05E+47 \\
$E_{\rm H}$ & 3.00E+49 & 1.11E+52 & 1.81E+50 & 1.87E+51 & 1.92E+50 & 3.20E+46 & 9.26E+45 & 1.34E+52 \\
$E_{\rm He}$ & --- & --- & 2.57E+48 & 9.64E+50 & 4.25E+50 & 7.61E+44 & --- & 1.39E+51 \\
$E_\nu$(nucl) & 2.21E+48 & 7.73E+50 & 1.24E+49 & 1.27E+50 & 1.34E+49 & 2.05E+45 & 2.45E+44 & 9.28E+50 \\
$E_\nu$(ther) & --- & 5.54E+43 & 1.00E+45 & 6.37E+47 & 6.75E+49 & 6.40E+48 & 1.33E+47 & 7.47E+49 \\
$\Delta \Omega$ & $-$7.71E+49 & 1.78E+49 & 3.06E+49 & $-$2.95E+49 & $-$2.93E+50 & $-$2.96E+49 & $-$5.67E+48 & $-$3.87E+50 \\
$\Delta U$ & 3.85E+49 & $-$8.32E+48 & $-$1.57E+49 & 1.57E+49 & 1.84E+50 & 2.15E+49 & 4.22E+48 & 2.42E+50 \\
$\Delta M$(X) & $-$2.59E$-$03 & $-$9.37E$-$01 & $-$1.51E$-$02 & $-$1.58E$-$01 & $-$1.62E$-$02 & $-$3.18E$-$06 & $-$7.57E$-$07 & $-$1.13E+00 \\
$\Delta M$(Y) & 1.61E$-$03 & 9.39E$-$01 & 1.31E$-$02 & $-$4.83E$-$01 & $-$2.58E$-$01 & 3.18E$-$06 & 7.57E$-$07 & 2.13E$-$01 \\
$\Delta M$(CNO) & 1.04E$-$03 & $-$1.74E$-$03 & 2.00E$-$03 & 6.28E$-$01 & 2.70E$-$01 & 2.14E$-$08 & --- & 8.99E$-$01 \\
$\Delta M$(Ne+Mg) & --- & --- & --- & 1.34E$-$02 & 5.48E$-$03 & --- & --- & 1.89E$-$02 \\
$\delta$(X) & 1.83E$-$04 & 1.85E$-$02 & 3.58E$-$03 & 7.74E$-$02 & 3.00E+00 & --- & --- & 3.10E+00 \\
$\delta$(Y) & 7.16E$-$05 & 7.21E$-$03 & 1.49E$-$03 & 3.40E$-$02 & 1.82E+00 & --- & --- & 1.86E+00 \\ 
$\delta$(CNO) & 3.60E$-$06 & 3.62E$-$04 & 7.16E$-$05 & 1.57E$-$03 & 6.89E$-$02 & --- & --- & 7.09E$-$02 \\
$\delta$(Ne+Mg) & 6.12E$-$07 & 6.16E$-$05 & 1.21E$-$05 & 2.66E$-$04 & 1.14E$-$02 & --- &--- & 1.17E$-$02 \\
\hline
\tablebreak 
\multicolumn{9}{l}{$7{\rm M_\odot}$} \\ \hline
Age & 0.47 & 46.8 & 47.1 & 52.7 & 53.02 & 60.7 & 6812 \\
Mass & 7.00 & 6.97 & 6.96 & 6.81 & 1.04 & 1.04 & 1.04 \\
log $L/L_\odot$ & 3.16 & 3.38 & 3.66 & 3.69 & 4.16 & 0.19 & $-$2.73 \\
$R$ & 12.4 & 4.46 & 64.5 & 97.6 & 304 & 0.00789 & 0.00746 \\
log $T_{\rm eff}$ & 4.14 & 4.29 & 4.00 & 3.73 & 3.57 & 4.70 & 3.94 \\
log $g_{\rm sup}$ & 3.90 & 4.06 & 3.05 & 1.64 & 0.59 & 8.66 & 8.71 & --- \\
\hline 
$E_\gamma$ & 8.39E+49 & 1.36E+52 & 1.76E+50 & 3.34E+51 & 5.80E+50 & 1.47E+48 & 1.54E+48 & 1.78E+52 \\
\phm{.....}  $E_{\rm NIR}$  &  5.76E+48 &  2.28E+50  & 4.44E+49  & 1.18E+51  & 3.71E+50 & 3.75E+44 & 2.98E+46 & 1.83E+51 \\
\phm{.....}  $E_{\rm Opt}$  &  2.44E+49 &  2.68E+51  & 7.34E+49  & 1.87E+51  & 2.08E+50 & 6.75E+45 & 2.31E+47 & 4.86E+51 \\
\phm{.....}  $E_{\rm  UV}$ &  5.36E+49  & 1.07E+52  &  5.84E+49  & 2.88E+50  & 1.70E+48 & 1.31E+48 & 1.28E+48 & 1.11E+52 \\
\phm{.....} $E_{\rm  X}$ & ---  & ---  & --- &  --- & --- & 1.48E+47 &  --- & 1.48E+47 \\
$E_{\rm H}$ & 3.62E+49 & 1.36E+52 & 1.91E+50 & 2.41E+51 & 1.82E+50 & 1.54E+45 & 1.64E+45 & 1.62E+52 \\
$E_{\rm He}$ & --- & --- & 3.49E+48 & 1.19E+51 & 3.39E+50 & 5.93E+44 & --- & 1.53E+51 \\
$E_\nu$(nucl)  &  2.55E+48 &  9.45E+50  & 1.31E+49  &  1.45E+50  & 1.26E+49  & 9.86E+44 & 4.43E+43 & 1.12E+51 \\
$E_\nu$(ther) & ---  & 6.25E+43 & 1.10E+45 & 7.02E+47 &  1.16E+50 & 6.34E+48 & 1.44E+47 & 1.25E+50 \\
$\Delta \Omega$ & $-$9.53E+49 & 2.09E+49 & 3.65E+49 & $-$3.80E+49 & $-$4.47E+50 & $-$3.37E+49 & $-$8.14E+48 & $-$5.67E+50 \\
$\Delta U$ & 4.76E+49 & $-$9.41E+48 & $-$1.89E+49 & 2.05E+49 & 2.64E+50 & 2.59E+49 & 6.47E+48 & 3.39E+50 \\
$\Delta M$(X)  & $-$3.31E$-$03 & $-$1.15E+00  & $-$1.62E$-$02 & $-$1.78E$-$01  & $-$1.60E$-$02 & $-$1.20E$-$06 & $-$9.50E$-$08 & $-$1.36E+00 \\
$\Delta  M$(Y) &  2.76E$-$03 &  1.15E+00  & 1.35E$-$02  & $-$6.2E$-$01  & $-$2.08E$-$01  & 1.20E$-$06 & 9.50E$-$08 & 3.38E$-$01 \\
$\Delta M$(CNO) & 1.24E$-$03 & $-$2.16E$-$03 & 2.50E$-$03 & 7.75E$-$01 & 2.16E$-$01 & 9.00E$-$07 & --- & 9.92E$-$01 \\
$\Delta M$(Ne+Mg) & --- & --- & 1.18E$-$4 & 1.67E$-$02 & 5.94E$-$03 & --- & --- & 2.27E$-$02 \\
$\delta$(X) & 2.47E$-$04 & 2.12E$-$02 & 4.69E$-$03 & 1.04E$-$01 & 3.43E+00 & --- & --- & 3.56E+00 \\
$\delta$(Y) & 9.64E$-$05 & 8.27E$-$03 & 1.97E$-$03 & 4.55E$-$02 & 2.23E+00 & --- & --- & 2.29E+00 \\
$\delta$(CNO) & 4.84E$-$06 & 4.15E$-$04 & 9.40E$-$05 & 2.11E$-$03 & 8.02E$-$02 & --- & --- & 8.29E$-$02 \\
$\delta$(Ne+Mg) & 8.24E$-$07 & 7.06E$-$05 & 1.59E$-$05 & 3.58E$-$04 & 1.34E$-$02 & --- &--- & 1.39E$-$02 \\
\hline
\tablebreak 
\multicolumn{9}{l}{$8{\rm M_\odot}$} \\ \hline
Age & 0.33 & 35.4 & 35.6 & 39.6 & 39.94 & 47.73 & 4811 & \\
Mass & 8.0 & 7.97 & 7.94 & 7.54 & 1.15 & 1.15 & 1.15 & \\
log $L/L_\odot$ & 3.37 & 3.58 & 3.86 & 3.88 & 4.17 & 0.14 & $-$2.56 & \\
$R$ & 14.8 & 4.82 & 79.3 & 136 & 300 & 0.00645 & 0.00619 & \\
log $T_{\rm eff}$ & 4.17 & 4.32 & 4.04 & 3.71 & 3.57 & 4.74 & 4.02 & \\
log $g_{\rm sup}$ & 3.88 & 4.05 & 3.08 & 1.43 & 0.50 & 8.88 & 8.92 & --- \\
\hline 
$E_\gamma$ & 9.54E+49 & 1.63E+52 & 1.82E+50 & 3.65E+51 & 5.95E+50 & 1.25E+48 & 1.57E+48 & 2.09E+52 \\
\phm{.....}   $E_{\rm NIR}$  & 6.46E+48  & 2.23E+50  & 4.42E+49  &  1.43E+51 & 3.59E+50 & 2.55E+44 & 1.90E+46 & 2.06E+51 \\
\phm{.....}   $E_{\rm Opt}$  & 2.55E+49  & 2.76E+51  & 7.03E+49  &  1.96E+51 & 2.34E+50 & 4.66E+45 & 1.73E+47 & 5.05E+51 \\
\phm{.....}  $E_{\rm  UV}$ &  6.34E+49  & 1.34E+52  &  6.71E+49  & 2.62E+50  & 2.14E+48 & 9.89E+47 & 1.38E+48 & 1.37E+52 \\
\phm{.....} $E_{\rm  X}$ & ---  & ---  & --- &  --- & ---  & 2.56E+47 &  --- & 2.46E+47 \\
$E_{\rm H}$ & 3.89E+49 & 1.64E+52 & 1.97E+50 & 2.18E+51 & 5.60E+49 & 7.32E+45 & 1.43E+43 & 1.88E+52 \\
$E_{\rm He}$ & --- & --- & 4.34E+48 & 1.43E+51 & 4.96E+50 & --- & --- & 1.93E+51 \\
$E_{\rm C}$ & --- & --- & --- & --- & 2.87E+50 & --- & --- & 2.87E+50 \\ 
$E_\nu$(nucl)  &  2.72E+48 &  1.13E+51  & 1.36E+49  &  1.49E+50  & 6.78E+48  & 4.68E+44 & 3.86E+41 & 1.30E+51 \\
$E_\nu$(ther)  &  7.04E+43 &  1.16E+45  & 8.08E+47  &  6.78E+48  & 4.58E+50  & 6.94E+48 & 1.61E+47 & 4.66E+50 \\
$\Delta \Omega$ & $-$1.13E+50 & 2.40E+49 & 4.15E+49 & $-$4.73E+49 & $-$7.49E+50 & $-$4.83E+49 & $-$1.17E+49 & $-$9.05E+50 \\
$\Delta U$ & 5.65E+49 & $-$1.07E+49 & $-$2.16E+49 & 2.56E+49 & 5.33E+50 & 4.01E+49 & 9.92E+49 & 5.65E+50 \\
$\Delta M$(X) & $-$3.59E$-$03 & $-$1.39E+00 & $-$1.66E$-$02 & $-$1.85E$-$01 & $-$4.39E$-$02 & $-$5.64E$-$07 & $-$9.40E$-$08 & $-$1.64E+00 \\
$\Delta M$(Y) & 2.23E$-$03 & 1.39E+00 & 1.36E$-$02 & $-$6.42E$-$01 & $-$2.54E$-$01 & 5.64E$-$07 & 9.40E$-$08 & 5.09E$-$01 \\
$\Delta M$(CNO) & 1.44E$-$03 & $-$2.14E$-$03 & 2.75E$-$03 & 8.09E$-$01 & $-$6.83E$-$02 & --- & --- & 7.43E$-$01 \\
$\Delta M$(Ne+Mg) & --- & --- & 2.40E$-$04 & 1.77E$-$02 & 3.09E$-$01 & --- & --- & 3.27E$-$01 \\
$\delta$(X) & 2.75E$-$04 & 2.41E$-$02 & 1.52E$-$02 & 1.21E$-$01 & 3.61E+00 & --- & --- & 3.93E+00 \\
$\delta$(Y) & 1.07E$-$04 & 9.40E$-$03 & 6.46E$-$03 & 5.63E$-$03 & 2.66E+00 & --- & --- & 2.79E+00 \\
$\delta$(CNO) & 5.40E$-$06 & 4.72E$-$04 & 3.05E$-$04 & 9.55E$-$04 & 8.82E$-$02 & --- & --- & 9.46E$-$02 \\
$\delta$(Ne+Mg) & 9.18E$-$07 & 8.03E$-$05 & 5.17E$-$05 & 9.55E$-$04 & 1.49E$-$02 & --- & --- & 1.60E$-$02 \\
\enddata
\end{deluxetable}

\begin{deluxetable}{lrrr}
\tabletypesize{\small}
\tablewidth{0pt}
\tablecolumns{4}
\tablecaption{Evolutionary  characteristics   of a $0.8{\rm M}_\odot$ star. 
\label{tab:evol0.8}}  
\tablehead{ \colhead{} & \colhead{PMS} &
\colhead{MS}   &  \colhead{RGB}} 
\startdata
\multicolumn{3}{l}{$0.8{\rm M_\odot}$} \\ \hline
Age & 73.2 & 25560 & 30211 \\
Mass & 0.8 & 0.798 & 0.709 \\
log $L/L_\odot$ & $-$0.45 & $-$0.40 & 0.77 \\
$R$ & 0.90 & 0.81 & 2.27 \\
log $T_{\rm eff}$ & 3.66 & 3.70 & 3.70 \\
log $g_{\rm sup}$ & 4.51 & 4.54 & 4.10 \\
\hline
$E_\gamma$ & 3.01E+48 & 1.22E+51 & 3.07E+51 \\
\phm{.....}   $E_{\rm NIR}$ & 1.32E+48 & 4.33E+50 & 1.85E+51 \\
\phm{.....}   $E_{\rm Opt}$ & 1.63E+48 & 7.42E+50 & 1.44E+51 \\
\phm{.....}   $E_{\rm UV}$ & 5.64E+46 & 4.65E+49 & 3.98E+49 \\
\phm{.....}   $E_{\rm X}$ & --- & --- & --- \\
$E_{\rm H}$ & 9.31E+47 & 1.22E+51 & 3.05E+51 \\
$E_{\rm He}$ & --- & --- & 1.73E+47 \\
$\Delta \Omega$ & $-$4.16E+48 & $-$2.40E+47 & $-$3.83E+49 \\
$\Delta U$ & 2.08E+48 & 1.33E+47 & 2.46E+49 \\
$E_\nu$(nucl) & 2.38E+46 & 2.99E+49 & 2.01E+50 \\
$E_\nu$(ther) & --- & 1.66E+43 & 2.82E+47 \\
$\Delta M$(X) & $-$1.31E$-$04 & $-$9.71E$-$02 & $-$2.40E$-$01 \\
$\Delta M$(Y) & 1.32E$-$04 & 9.70E$-$02 & 2.40E$-$01 \\
$\Delta M$(CNO) & 8.86E$-$07 & 1.40E$-$04 & $-$2.55E$-$04 \\
$\Delta M$(Ne+Mg) & --- & --- & --- \\
$\delta$(X) & --- & 1.16E$-$03 & 6.22E$-$02 \\
$\delta$(Y) & --- & 4.52E$-$04 & 2.57E$-$02 \\
$\delta$(CNO) & --- & 2.27E$-$05 & 1.24E$-$03 \\
$\delta$(Ne+Mg) & --- & 3.86E$-$06 & 2.11E$-$04 \\
\enddata
\end{deluxetable}

\begin{deluxetable}{lcccccccc}  
\tablecolumns{7}  
\tablewidth{0pc}  
\tablecaption{Mean values of $R$ and $K$.}
\tablehead{
\colhead{$M$} & \colhead{$R$(MS)} & \colhead{$K$(MS)} & \colhead{$R$(RGB)} &\colhead{$K$(RGB)}&\colhead{$R$(He-core)}&\colhead{$K$(He-core)}&\colhead{$R$(WD)}&\colhead{$K$(WD)}}
\startdata
0.8 & 0.806& 1.382& 2.265& 2.610& -& -& -& -\cr
1   & 1.000& 1.594& 2.603& 2.601& 9.743& 59.81& 0.0138& 0.927\cr
1.5 & 1.790& 2.300& 10.02& 5.587& 10.75& 27.57& 0.0130& 0.927\cr
1.8 & 2.108& 2.328& 15.62& 7.385& 10.13& 13.39& 0.0131& 0.925\cr 
2   & 2.239& 2.291& 8.158& 4.072& 8.276& 6.981& 0.0128& 0.925\cr 
3   & 2.844& 2.148& 14.55& 4.795& 14.71& 7.758& 0.0125& 0.925\cr
4   & 3.302& 2.036& 22.16& 4.974& 30.62& 13.36& 0.0102& 0.942\cr
5   & 3.709& 1.954& 33.83& 5.006& 52.49& 20.76& 0.0096& 0.949\cr
6   & 4.095& 1.900& 48.58& 5.097& 71.80& 23.57& 0.0085& 0.967\cr
7 & 4.463& 1.858& 64.50& 5.151& 97.65& 26.59& 0.0074 & 0.993 \cr 
8 & 4.821& 1.825& 79.34& 5.123& 135.82& 35.69& 0.0061 & 1.033\cr
\enddata
\label{table:counts}
\end{deluxetable}

\end{document}